\documentclass[twoside,11pt]{article}

\usepackage{amsmath}
\usepackage{graphicx,psfrag,epsf}
\graphicspath{{./fig/}}
\usepackage{enumerate}
\usepackage{url} 
\usepackage{fancyvrb}
\usepackage{microtype}
\usepackage[dvipsnames]{xcolor}
\definecolor{matblue}{rgb}{0,0.447,0.741}    
\definecolor{matred}{rgb}{0.85,0.325,0.098}
\definecolor{matyellow}{rgb}{0.929,0.694,0.125}
\definecolor{matpurple}{rgb}{0.494,0.184,0.556}
\definecolor{matgreen}{rgb}{0.466,0.674,0.188}
\definecolor{matcyan}{rgb}{0.301,0.745,0.933}
\definecolor{matdarkred}{rgb}{0.635,0.078,0.184}
\usepackage{subcaption}
\usepackage[preprint]{jmlr2e}
\usepackage[boxed]{algorithm2e}
\SetAlgoInsideSkip{medskip}
\SetAlCapSkip{11pt}
\SetAlCapSty{textrm}
\SetAlgoCaptionLayout{raggedright}

\DeclareMathOperator{\N}{\mathrm{N}}
\DeclareMathOperator{\E}{\mathrm{E}}
\DeclareMathOperator{\exponential}{\mathrm{exponential}}
\DeclareMathOperator{\Var}{\mathrm{Var}}
\DeclareMathOperator{\KL}{\mathrm{KL}}
\def\effmcmc{\mathrm{eff,MCMC}}

\hypersetup{
  pdftitle={Pareto Smoothed Importance Sampling},
  pdfauthor={Aki Vehtari,
    Daniel Simpson,
    Andrew Gelman,
    Yuling Yao, and
    Jonah Gabry},
  pdfkeywords={importance sampling,
    Monte Carlo,
    Bayesian computation,
    diagnostics,
    very good importance sampling},
  hidelinks
}

\usepackage{lastpage}
\jmlrheading{25}{2024}{1-\pageref{LastPage}}{7/19; Revised
8/22}{3/24}{19-556}{Aki Vehtari, Daniel Simpson, Andrew Gelman, Yuling Yao, and Jonah Gabry}
\ShortHeadings{Pareto Smoothed Importance Sampling}{Vehtari, Simpson, Gelman, Yao, and Gabry}
\firstpageno{1}

\begin{document}

\title{Pareto Smoothed Importance Sampling}

\author{%
\name Aki Vehtari \email aki.vehtari@aalto.fi \\
\addr Department of Computer Science\\
Aalto University\\
\AND
\name Daniel Simpson \email dan@normalcomputing.ai \\
\addr Normal Computing\\
\AND
\name Andrew Gelman \email gelman@stat.columbia.edu \\
\addr Departments of Statistics and Political Science\\
Columbia University\\
\AND
\name Yuling Yao \email yyao@flatironinstitute.org \\
\addr Center for Computational Mathematics\\
Flatiron Institute\\
\AND
\name Jonah Gabry \email jgabry@gmail.com \\
\addr Department of Statistics\\
Columbia University\\
}

\editor{Vikash Mansinghka}

\maketitle

\begin{abstract}
Importance weighting is a general way to adjust Monte Carlo integration to
account for draws from the wrong distribution, but the resulting estimate can be
highly variable when the importance ratios have a heavy right tail. This routinely occurs
when there are aspects of the target distribution that are not well captured by
the approximating distribution, in which case more stable estimates can be
obtained by modifying extreme importance ratios. We present a new method for
stabilizing importance weights using a generalized Pareto distribution fit to
the upper tail of the distribution of the simulated importance ratios. The
method, which empirically performs better than existing methods for stabilizing
importance sampling estimates, includes stabilized effective sample size
estimates, Monte Carlo error estimates, and convergence diagnostics.
The presented Pareto $\hat{k}$ finite sample convergence rate diagnostic
is useful for any Monte Carlo estimator.
\end{abstract}


\section{Introduction}

Importance sampling is a simple modification to the Monte Carlo method for
computing expectations that is useful when there is an auxiliary distribution
$g(\theta)$ that is easier to directly sample from than the target distribution
$p(\theta)$, which may be only known up to a proportionality constant
\citep{Hammersley:1964}. The
starting point is the simple Monte Carlo estimate for the expectation of
a function $h$,
$$
I_h
  = \E_p(h)
  = \int h(\theta) p(\theta)\,d\theta
  \approx \frac{1}{S} \sum_{s=1}^S h(\theta_s),
$$
which requires exact draws $\theta_s, \ s=1,\dots,S$ from $p(\theta)$.
The self-normalized importance sampling estimate for the same expectation is
\begin{equation} \label{eqn:SNIS}
\frac{ \sum_{s=1}^S r_s h(\theta_s)}{\sum_{s=1}^S r_s},
\qquad r_s = r(\theta_s) = \frac{p(\theta_s)}{g(\theta_s)},
\end{equation}
which only requires draws $\theta_s$ from a proposal distribution $g(\theta)$.

The success of the importance sampling estimator depends on the
distribution of the importance ratios $r_s$ and of $r_s h(\theta_s)$. When the proposal distribution is a
poor approximation to the target distribution, the distribution of importance
ratios can have a heavy right tail. This can lead to unstable importance
weighted estimates, sometimes with infinite variance.

Textbook examples of poorly performing importance samplers occur in low
dimensions when the proposal distribution has lighter tails than the target, but
it would be a mistake to assume that heavy-tailed proposals will always stabilize
importance samplers.  This intuition is particularly misplaced in high
dimensions, where importance sampling can fail even when the ratios have finite
variance. \citet[Section 29.2]{mackay2003information} provides an example of
what goes wrong in high dimensions:  the ratios $r_s$ can vary by
several orders of magnitude and the estimator \eqref{eqn:SNIS} will be dominated
by a few draws. Hence, even if the approximating distribution is chosen so that
the importance ratios are bounded so that \eqref{eqn:SNIS} has finite variance,
the bound can be so high and the variance so large that the behavior of the
self-normalized importance sampling estimator is practically indistinguishable
from an estimator with infinite variance. This suggests that if we want an
importance sampling method that works in high dimensions, we need to move beyond
being satisfied with estimates that have finite variance and find methods
with built-in error assessment.

In this paper, we
\begin{itemize}
\item propose Pareto smoothed importance sampling (PSIS), a method for stabilizing importance sampling estimates;
\item show that PSIS has the usual
properties for a well-behaved self-normalized importance sampling estimator such as consistency and finite variance;
\item propose a simple numerical Pareto $\hat{k}$ diagnostic for finite sample convergence rate suitable for any Monte Carlo estimator;
\item demonstrate the behavior of PSIS and $\hat{k}$ in several low and high-dimensional examples.
\end{itemize}
This paper focuses on self-normalized importance sampling, but Pareto $\hat{k}$ diagnostic and Pareto smoothing can be used also for ordinary importance sampling or any Monte Carlo estimate, as demonstrated in some of the references discussed in Section \ref{s:discussion}.
Beyond the examples in the latter part of
this paper, PSIS forms the basis for the widely-used \textsc{loo} R package for
stable, high-dimensional leave-one-out cross-validation
\citep{Vehtari+Gelman+Gabry:2017,loo-software:2024}.
This package has been downloaded more than three million times.
The PSIS and Pareto $\hat{k}$ diagnostic have been also implemented, for example, in the \textsc{posterior} R package \citep{posterior-software:2023}, the \textsc{ArviZ.py} and \textsc{Arviz.jl} packages for Python and Julia \citep{Kumar+etal:2019:ArviZ}, and the 
\textsc{Pyro} Python package \citep{bingham2019pyro}.

\section{Stabilizing Importance Sampling Estimates by Modifying the Ratios}

The stability of self-normalized importance sampling can be improved by
directly modifying the computed ratios. For notational convenience, we rewrite
these importance samplers as,
\begin{equation} \label{eqn:SNIS2}
\int h(\theta) p(\theta)\,d\theta \approx \frac{ \sum_{s=1}^S w_s h(\theta_s)}{\sum_{s=1}^S w_s},
\end{equation}
where $w_s=r_s$ would recover the standard self-normalized importance sampler.

\citet{Ionides:2008} showed that the truncation rule,
\begin{equation}\label{min}
  w_s = \min{\left(r_s,\sqrt{S}\bar{r}\right)},
\end{equation}
where $\bar{r}$ is the average of the original $S$ importance ratios, is
simulation-consistent with finite variance. 
The critical advantage
conveyed by the truncation is that these properties now extend to problems that
only have integrable ratios: we get finite variance 
under the
assumption that $\mathbb{E}(|r_s|)< \infty$ instead of under the stronger
condition $\mathbb{E}(|r_s|^2)< \infty$ which is needed for the unmodified
importance sampling estimator \eqref{eqn:SNIS} to have finite variance.

This simple modification to the raw importance ratio greatly extends the range
of problems for which the estimator has finite variance. 
Unfortunately, while the truncation can improve the stability of the
weights, our experiments show that the simple weight modification scheme
can be too severe, leading to larger than necessary finite sample bias.

\subsection{Modeling the Tail of the Importance Ratios}

In this paper, we propose a new scheme for modifying the extreme importance
ratios that adapts more readily to the problem under consideration than the
universal truncation rule of \citet{Ionides:2008}.

To motivate the new scheme, we begin by noting that the success of plain
importance sampling depends on how many moments $r(\theta)$ and
$h(\theta)r(\theta)$ possess, with the estimator having finite variance if
at least two moments exist.
This suggests that using information about the
distribution of $r_s | r_s\!>\!u$, for some threshold $u \rightarrow \infty$ as
$S\rightarrow \infty$, should allow us to improve the quality of our importance
sampling estimators.

\citet{Pickands:1975} and \citet{Balkema+Haan:1974} proved, under commonly satisfied conditions, that as the
sample size increases, the distribution of  $r_s | r_s\!>\!u$ is well
approximated by the three-parameter generalized Pareto distribution,
\begin{align}\label{pareto}
  p(y \mid u,\sigma,k)=
  \begin{cases}
    \frac{1}{\sigma}\left(1+k\left(\frac{y-u}{\sigma}\right)\right)^{-\frac{1}{k}-1}, & k\neq 0 \\
    \frac{1}{\sigma}\exp\left(\frac{y-u}{\sigma}\right), & k = 0,
  \end{cases}
\end{align}
where $u$ is a lower bound parameter, $y$ is restricted to the range
$(u,\infty)$, $\sigma$ is a non-negative scale parameter, and $k$ is an
unconstrained shape parameter.
The generalized Pareto distribution has $1/k$ finite fractional moments
when $k>0$, and thus we can infer the number of existing fractional moments of the weight
distribution by focusing on $k$.

To estimate the parameters in the generalized Pareto distribution, we use the
$M$ largest importance ratios, where
$$
M = \lfloor \min(0.2 S, 3\sqrt{S})\rfloor,
$$
where $\lfloor \cdot \rfloor$ denotes the floor function.
Restricting the tail modeling to a subset of the largest importance ratios
implicitly defines a value of  $u$ in the Pareto distribution. The above rule
for $M$ was made based on extensive computational experiments, related suggestions in the literature \citep{Scarrot+MacDonald:2012}, and in line with the
requirements for consistent estimation \citep{Pickands:1975}.
The minimum tail proportion $0.2$ reduces the possible bias for small $S$, with the square root rule providing stability when $S$ is high.
In practice, we
have found the method to be insensitive to the exact form of $M$; for
instance, using $M=0.2 S$ in all cases was suggested by
\citet{Vehtari+Gelman+Gabry:2017}, and it worked well even though it is not
asymptotically correct.  We tested several other methods reviewed by
\citet{Scarrot+MacDonald:2012} for selecting $u$
directly, but found them to have higher variability than this simple
heuristic, leading to increased variance in estimating $k$.

The scale and shape parameters, $\sigma$ and $k$, can be estimated using the
highly efficient, low bias approximate Bayesian method of \citet{Zhang+Stephens:2009}, which we briefly review in Appendix \ref{sec:marginal_k}. We choose this
approach due to its efficiency and its ability to be used automatically without
human intervention. In Appendix \ref{sec:regularization_of_k} we describe
additional practical regularization of the estimate $\hat{k}$, which helps to reduce the
variance of the estimate when $S$ is small, without affecting its asymptotic
properties.

Although in most cases we cannot verify that the distribution of $r_s$ lies in
the domain of attraction for an extremal distribution, we will use this as a
working assumption for building our method. \citet{Pickands:1975} notes that
``most `textbook' continuous distribution functions'' lie in
the domain of attraction of some extremal distribution function.  For finite
$S$, we could alternatively interpret $\hat{k}$ as saying that the sample
$(r(\theta_s))_{s=1}^S$ is not statistically distinguishable from a sample of
size $S$ from a distribution with tail index $\hat{k}$.
\begin{algorithm}[t]
\KwIn{Raw importance ratios $r_s$, $s=1,\ldots,S$, ordered from lowest to highest}
\KwOut{PSIS-smoothed importance weights $w_s$, $s=1,\ldots,S$}
\BlankLine
Set $M = \lfloor\min(0.2 S, 3\sqrt{S})\rfloor$\;
Set $w_s =  r_s$, $s = 1,\ldots,S-M$\;
Set $\hat{u}=r_{S-M}$\;
Estimate $(\hat{k},\hat{\sigma})$ in the generalized Pareto distribution with cutpoint $\hat{u}$ from
the $M$ largest importance ratios, using the algorithm of
\citet{Zhang+Stephens:2009} with the additional prior described in
Appendix~\ref{sec:regularization_of_k}\;

Set $ w_{S-M+z} =\min\left(F^{-1}\left(\frac{z-1/2}{M}\right),\max_s(r_s)\right)$,
for each $z=1,\ldots,M$\;

If the estimated shape parameter $\hat{k}$ exceeds $\min\left(1-1/\log_{10}(S), 0.7\right)$, report a warning that
the resulting importance sampling estimates are likely to be unstable or have high bias. For $S > 2000$ this threshold is $0.7$.
\caption{PSIS procedure for computing importance weights.\hfill}
\label{algo:PSIS}
\end{algorithm}

\subsection{Our Proposal: Pareto Smoothed Importance Sampling}\label{s:pareto}

We propose a new method to stabilize the importance weights by replacing the $M$
largest weights above the threshold $u$ by a set of well-spaced values that are
consistent with the tails of the importance distribution,
\begin{align*}
w_{S-M+z}
 &= F^{-1}\left(\frac{z-1/2}{M}\right)  \\
 &= u + \frac{\sigma}{k}\left(\left(1-\frac{z-1/2}{M}\right)^{-k} - 1\right),
\end{align*}
where $z=1,\dots,M$, and $F^{-1}$ is the inverse-CDF of the generalized Pareto
distribution fitted to the $M$ largest importance ratios.
The inverse transformation corresponds to a fast approximation
of the expected order statistics \citep{Blom:1958}, with increasing accuracy as
$M$ increases. Expected order statistics provide low bias and reduced variance
compared to the original ratios or ordered random draws from the
distribution. 

We show that replacing the largest ratios with the expected order statistics changes PSIS to have finite variance and an error distribution converging to normal, when $k\in (0.5,1)$.  Section \ref{sec:k_hat_big} shows how the PSIS error distribution can be modelled with a distribution of mean of truncated Pareto variables with truncation corresponding to the largest expected order statistics.  Section \ref{sec:asymptotic} provides finer details proving that the PSIS
estimator, summarized in  Algorithm \ref{algo:PSIS}, is simulation-consistent and has finite variance under relatively light conditions. 
The final step in the algorithm warns the user if the estimated shape
parameter $\hat{k}$ in the generalized Pareto distribution is larger than $\min\left(1-1/\log_{10}(S), 0.7\right)$. This is
an automatic stability check that we justify
in Section \ref{sec:k_hat_big}.

When Pareto $\hat{k}$ is less than $\min\left(1-1/\log_{10}(S), 0.7\right)$, we recommend estimating the Monte Carlo standard error (MCSE) of PSIS using the estimator of \citet[][Ch.\ 9]{Owen:2013} \citep[see also][]{Elvira+etal:2022:rethinking_ESS} for self-normalized importance sampling,
\begin{eqnarray}\label{eq:VarISh}
\widehat{\widetilde{\Var}}(\hat{I}_h^S)
 & \approx& \sum_{s=1}^S \tilde{w}_s^2(h(\theta_s)-\tilde{\mu})^2, \mbox{ \ \  for independent draws} \\\label{eq:VarISh_MCMC}
&& \sum_{s=1}^S \tilde{w}_s^2(h(\theta_s)-\tilde{\mu})^2/R_{\effmcmc}, \mbox{ \ \ for MCMC draws},
\end{eqnarray}
where $\tilde{w}_s = w_s / \sum_{s'=1}^Sw_{s'}$ are normalized weights,
$\tilde{\mu}=\sum_{s=1}^S \tilde{w}_sh(\theta_s)$, and $R_{\effmcmc}=S_{\effmcmc}/S$ is the relative efficiency of MCMC, and the
effective sample size $S_{\effmcmc}$ for $h(\theta)$ is computed using the
split-chain method \citep{Vehtari+etal:2021:Rhat}.

The corresponding effective sample size estimate is
\begin{equation}\label{eq:ESSh}
\mathrm{ESS}_h \approx \frac{\frac{1}{S}\sum_{s=1}^S \left(h(\theta_s)-\frac{1}{S}\sum_{s=1}^S h(\theta_s)\right)^2}{\widehat{\widetilde{\Var}}(\hat{I}_h^S)}.
\end{equation}
The effective sample size for the normalization term (zeroth moment) is 
\begin{equation}\label{eq:ESS}
\mathrm{ESS} \approx 1/{\sum_{s=1}^S \tilde{w}_s^2}.
\end{equation}
Expression (\ref{eq:ESS}) is also the generic effective sample size estimate proposed by \citet{Kong:1992} \citep[see also][]{Kong+Liu+Wong:1994,Owen:2013,Elvira+etal:2022:rethinking_ESS}, which doesn't depend on $h$.
\citet{Chatterjee+Diaconis:2018} warn against using a variance estimator
as an importance sampling diagnostic, as it depends on
  the accuracy of an estimate obtained by importance sampling itself,
  and they recommend using a ``diagnostic that is not itself an importance
  sampling estimate of any quantity.'' Pareto $\hat{k}$ is such a
  diagnostic, and  we
validate these approximations using the examples in Sections \ref{sec:k_hat_big} and \ref{sec:practical}, and Appendix \ref{sec:examples},
demonstrating good behavior when the estimated shape
parameter in the generalized Pareto distribution is less than $\min\left(1-1/\log_{10}(S), 0.7\right)$.

The following example shows how PSIS compares with simple importance sampling
(IS) and truncated importance sampling (TIS) for a simple one-dimensional example (with repeated simulations).  Further
simulated examples in low and moderate dimensions can be found in Section \ref{sec:k_hat_big} and Appendix
\ref{sec:examples}.

\begin{example}
Consider the following one-dimensional example where the target distribution is
$\exponential(1)$ and the proposal distributions are
$\exponential(\theta)$ for $\theta \in (1.3,1.5,2,3,4,10)$, with the
exponential distribution parameterized by the rate (inverse mean) parameter $\theta$.
The importance ratios will have infinite variance whenever $\theta>2$ \citep{Robert+Casella:2004}.

\begin{figure}[t]
  \begin{center}
    \includegraphics[width=\textwidth]{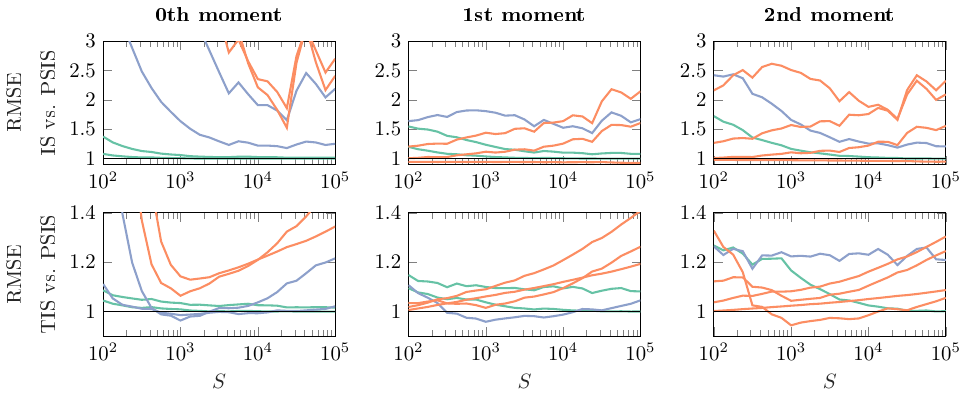}
  \end{center}
   \vspace{-0.75\baselineskip}
  \caption{For Example 1, the ratio of root mean squared error (RMSE, computed from 1000 simulations) between importance
  sampling (IS), truncated importance sampling (TIS), and Pareto smoothed
  importance sampling (PSIS) for zeroth, first $(h(\theta) = \theta)$, and second
  $(h(\theta) = \theta^2)$ moments. A ratio greater than $1$ means PSIS had lower RMSE
  than its competitor. The target distribution is $\exponential(1)$ and
  the proposal distribution is exponential with rate (inverse mean) parameter
  $\theta \in (1.3,1.5,2,3,4,10)$. The lines are colored according to the
  $\hat{k}$ (for zeroth moment) and $\hat{k}_h$ (for first and second moment) estimates. {\color{matgreen}Green lines have $\hat{k} < 0.5$}, {\color{matblue}blue lines have
  $\hat{k} \in (0.5, 0.7)$}, and {\color{matred}red lines have $\hat{k} > 0.7$}.
  }
  \label{fig:exptest_relrmse}
\end{figure}
Figure \ref{fig:exptest_relrmse} shows the ratio of root mean square errors
(RMSEs) computed from 1000 simulations, comparing ordinary (IS), truncated (TIS), and Pareto smoothed importance sampling (PSIS).
In all cases a ratio larger than $1$ corresponds to PSIS having the lower RMSE.
It is clear that PSIS is always better than straight IS. On the other hand, PSIS
is slightly worse than TIS with intermediate sample sizes for $\theta = 2$,
which corresponds to $\hat{k} \approx 0.5$.  A likely reason for this is that in
this case the truncation rule in TIS is perfectly calibrated for the weight
distribution and hence it is the best possible scenario for TIS. Deviations from
this scenario lead to PSIS performing better. This is the only example we found
where TIS outperformed PSIS.  Figure \ref{fig:exptest} in Appendix
\ref{sec:examples} shows the distribution of the weights and estimates over
several runs for this example when $\theta=3$. From this, we see that PSIS
yielded smaller bias but slightly larger variance than TIS.  We further explore
this example in Appendix \ref{sec:exponential_test}.
\end{example}

\section{Using $\hat{k}$ as a Diagnostic} \label{sec:k_hat_big}

As part of the PSIS procedure, we estimate the shape parameter $k$ of the
limiting generalized Pareto distribution for the upper tail of the importance ratios. 
A simple option would be to use this estimate $\hat{k}$ and trust
importance sampling as reliable if $\hat{k} < 0.5$, as it
would indicate that the variance is finite and the central limit theorem holds.
Our theory shows, however, that PSIS leads to valid and well-behaved importance
sampling routines for any integrable density. This suggests that requiring the
estimated shape parameter be less than $0.5$ would be unnecessarily stringent.

Occasionally, we will use an $h$-specific tail estimate $\hat{k}_h$ for the tails of the distribution $h(\theta)r(\theta)$, $\theta \sim g(\theta)$.
This can be
useful when $h(\theta)$ is unbounded, or goes to zero, as it is possible if $h$
grows fast enough (or goes to zero fast enough) relative to $r(\theta)$ that the
tail behavior of $r(\theta)h(\theta)$ can be qualitatively different from the tail
behavior of $r(\theta)$.  As $h(\theta)$ can be negative for some $\theta$ (e.g., when $h(\theta)=\theta$), we estimate $\hat{k}$ for both left and right tails of $h(\theta)r(\theta)$ and set $\hat{k}_h$ to that $\hat{k}$ which is bigger.
Although in this paper we focus on using $\hat{k}$ as a diagnostic for
PSIS, an $h$-specific tail estimate $\hat{k}_h$ is useful also when using
regular or quasi Monte Carlo (i.e., $r(\theta)=1$).
We validate the $h$-specific $\hat{k}_h$ in Appendix
\ref{sec:examples}.

\subsection{PSIS Is Reliable When $\hat{k} < 0.7$}
\label{sec:reliable_psis}

Extensive experiments show that for typical moderate sample sizes used in practice\footnote{The approximate threshold is $S>2000$, and for example, Stan  default is $S=4000$ \citep{Stan:2017}.},
PSIS gives reliable (that is, low bias, low
variance) estimators of $I_h$ when $\hat{k}_h <0.7$. This can be interpreted as
the estimate being reliable when the sample used to compute the PSIS estimate is
indistinguishable from one that comes from a density with more than
$0.7^{-1} \approx 1.4$ finite fractional moments.  When $\hat{k}_h>0.7$ (less than $1.4$ finite fractional moments) the convergence of the
estimator becomes too slow to be useful in many practical cases. We provide theoretical support of this threshold. We also provide a sample-size-specific threshold that is useful with small sample size ($S<2000$).

In this section, we distinguish between the true but unknown  $k$   and our 
finite sample estimate $\hat{k}$ as a way to understand the differences between the
asymptotic behavior of PSIS and its finite sample behavior. 
We first show theoretically that the computational complexity
of importance sampling has a meaningful qualitative change around $k=0.7$.  The finite sample estimate $\hat{k}$ is a useful indicator
of the practical pre-asymptotic convergence rate of PSIS even when the distribution
of ratios is bounded and has finite variance. In particular, we show that $\hat{k}$ can identify poorly
behaved but finite variance proposals in high dimensions.

\subsection{Pareto Means and Central Limit Theorems}
\label{sec:Pareto_GCLT}

Section \ref{sec:asymptotic} proves asymptotic consistency and finite variance. In this section we use various large sample results to characterize finite sample behavior of IS, TIS, and PSIS. There is no strict definition of large sample in each case, but the theory is able to explain many empirical results shown by our experiments.

We analyse the properties of IS, TIS, and PSIS using the generalized central limit theorem and distribution of sum of non-truncated and truncated Pareto distributed variables. In addition we use distribution of sum of truncated Pareto distributed variables separately to explain why $\hat{k}>0$ is a useful pre-asymptotic finite sample diagnostic when the bound of the bounded ratios is large.

\subsubsection{Generalized Central Limit Theorem}
The distribution of the mean of i.i.d. variables from a distribution with Pareto (power law) tails can be approximated with a stable distribution \citep[e.g.,][p. 62]{Uchaikin+Zolotarev:1999}.\footnote{\citet{Uchaikin+Zolotarev:1999} discuss sums, as these are well defined also when $k>1$, but we show the equations for means as the mean is our quantity of interest and we focus on $k<1$.} If the tail index $k<0.5$, the central limit theorem holds and the stable distribution is normal, scaling as $S^{-1/2}$ (as $S\rightarrow \infty$). For $k=0.5$ that stable distribution is a normal distribution that scales as $(S/\log(S))^{-1/2}$, and for $k\in (0.5,1)$ the stable distribution has stability parameter $1/k$ and scales as $S^{(k-1)}$ (as $S\rightarrow \infty$). The CLT and GCLT results for $k\neq 1/2$ are usually mentioned for $S\rightarrow \infty$, while the result for $k=0.5$ is reported conditional on $S$. We also sometimes refer to these well known results as is, but for finite $S$ the transition from $k<0.5$ via $k=0.5$ to $k>0.5$ is smooth. We eventually provide also an approximate smooth scaling rate for PSIS conditional on finite $S$.

We first consider just the estimation of the normalization term, that is the expectation of the ratios, which is already useful as this is what is needed, for example, in fast leave-one-out cross-validation \citep{Vehtari+Gelman+Gabry:2017}. To further simplify, we first consider just the tail part, which we assume is well approximated by Pareto distribution.

\subsubsection{Pareto Means and IS L1 Deviation}
  For $k\in (0.5,1)$, the explicit expressions for the densities of stable distributions are unknown, and numerical algorithms are needed.
\citet[][Eq.\ 28]{Zaliapin+etal:2005} provide a closed form approximation of quantile $q>0.95$ for the sum of Pareto-distributed variables.\footnote{The \citet{Zaliapin+etal:2005} preprint has a typo, which in the published version has been fixed by redefining the meaning of $q$ just for this equation, but we stick with the more natural definition of $q$ and rewrite the equation.} To get the corresponding approximation for the mean, we simply divide by $S$ to get
\begin{align}\label{eq:Pareto_mean_quantile}
S^{(k-1)} (1-q)^{-k} + \mu,
\end{align}
where $\mu$ is the expected mean.
As the distribution of ratios are right skewed, the distribution of mean will be also right skewed, and thus the absolute L1 deviation for the mean will be less than $S^{(k-1)} (1-q)^{-k}$ with probability $q$.

Thus to control the L1 deviation when $k$ increases beyond 0.5, $S$ needs to grow proportionally to
\begin{equation}\label{eq:L1_minimum_ESS_IS}
(c_1(1-q)^{-k})^{1/(1-k)},
\end{equation}
where $c_1$ includes also the effect of the scale of the distribution in more general case.
\citet{Zaliapin+etal:2005} provide more accurate numerical approximations of quantiles of the Pareto sum distribution (that can be used also for quantiles of mean distribution), but the accuracy of \eqref{eq:Pareto_mean_quantile} and \eqref{eq:L1_minimum_ESS_IS} is sufficient for our purposes when examining the L1 deviation of IS with raw ratios.

\subsubsection{Truncated Pareto Means and PSIS RMSE}
\label{sec:PSIS_RMSE}

PSIS replaces the largest ratios with their expected order statistics.
The  mean of the expected order statistics is not the same
as the expectation of mean of raw ratios. With thin-tailed distributions
the difference is usually small, but for thick-tailed distributions we
need to take the difference into account. The effect of replacing the
largest ratios with the expected order statistics is similar to
truncating with the largest expected order statistic.

Truncation makes PSIS have finite
variance (see Sections \ref{s:pareto} and \ref{sec:asymptotic}) with the
price of introducing bias. We can derive analytically the mean and
variance of the truncated Pareto mean; see Appendix \ref{sec:truncated_mean} and 
Eqs.\ (60) and (61) of \citet{Zaliapin+etal:2005}.  The
generalized central limit theorem states that the largest expected
order statistic scales as $S^k$ 
\citep[e.g.,][p.\ 138]{Bouchaud+Georges:1990}. By substituting $S^k$ in the
truncated mean and variance equations (see Appendix
\ref{sec:truncated_mean}), we see that standard deviation of PSIS
scales approximately as $S^{-1/2}$ when
$k < 0.5 -0.5/\log_{10}(S)$, as $(S\log(S))^{-1/2}$ when $k=0.5$,
and as $S^{k-1}$ when $k\in (0.5 + 0.5/\log_{10}(S), 1)$.  The bias of
PSIS is negligible when $k \leq 0.5$, and the bias scales approximately as
$S^{k-1}$ when $k\in (0.5+0.5/\log_{10}(S),  1)$. The scaling between the above-mentioned ranges and $k=1/2$ changes smoothly. When $k\in (0.5 +0.5/\log_{10}(S), 1)$, to control PSIS RMSE, $S$ needs to scale proportionally to
\begin{align}\label{eq:RMSE_minimum_ESS_PSIS}
c_2^{1/(1-k)}.
\end{align}
Based on the numerical experiments, $1/c_2$ matches  the target RMSE. Using a rule of 10\%, 
we get a rule of thumb, $S=10^{1/(1-k)}$. 
We can derive from the estimated minimum required sample size also an approximate ESS given $k$ as
\begin{align}\label{eq:ESS_given_k}
\mathrm{ESS}_k \approx S/10^{k/(1-k)}.
\end{align}
This is an optimistic estimate as it doesn't take into account the
variance of the weights, but it is useful to set expectations on sample size given observed $\hat{k}$. Although Eqs.\ \eqref{eq:RMSE_minimum_ESS_PSIS} and \eqref{eq:ESS_given_k} are derived for $k\in(0.5+0.5/\log_{10}(S), 1)$, they turn out to be surprisingly good approximations also for $k \leq 0.5+0.5/\log_{10}(S))$, as the scale of the error distribution depends on $k$.

Eqs.\ \eqref{eq:RMSE_minimum_ESS_PSIS} and \eqref{eq:ESS_given_k}
are meant only to give guidance on the order of magnitude for the required sample size.  When the sample size is sufficient and $\hat{k}<0.7$, a more accurate estimate of RMSE can be obtained using the variance (Eq.\ \ref{eq:VarISh}), and then given the convergence rate estimate discussed in Section \ref{sec:rate}, a more  accurate estimate of the required sample size to reach the required RMSE can be obtained. Especially when $\hat{k} < 0.5-0.5/\log_{10}(S)$ (see Section \ref{sec:rate}), we may assume the variance based RMSE and ESS estimates to be much more accurate than the above approximations depending only on $k$.

Based on numerical analysis of truncated Pareto means, when $k<0.7$ the variance dominates the PSIS RMSE. This explains why the variance based MCSE estimate (Eq.\ \ref{eq:VarISh}) works well for PSIS also when $k>0.5$ (as demonstrated by the examples). Due to the Pareto smoothing of $M$ largest weights, the PSIS RMSE is smaller than if only plain truncation at high upper quantile was used. When $k>0.7$, independently of $S$ the effect of bias dominates and the variance based MCSE starts to fail (also demonstrated by the examples). Although the accuracy can be improved by increasing $S$, diagnosing the accuracy is more difficult. This provides additional justification for the $0.7$ threshold. When the variance dominates the RMSE, we can also directly use the generalized CLT truncated variance result to justify the scaling of RMSE \citep[e.g.,][p.\ 138]{Bouchaud+Georges:1990}\footnote{\citet{Uchaikin+Zolotarev:1999} cite \citet{Bouchaud+Georges:1990}, but have a typo in their equation on p.\ 64.}.

TIS truncates at $\bar{r}S^{1/2}$, which reduces variance but increases bias. Due to a high bias of TIS, the variance based MCSE estimate for TIS starts to fail when $k>0.5$; see results in Appendix \ref{sec:exponential_test}. Eventually for larger $k$, the dominating bias makes the TIS RMSE scale as $S^{0.5-0.5/k}$ (see Appendix \ref{sec:truncated_mean}); that is, TIS RMSE increases much faster than PSIS RMSE when $k>0.5$.

RMSE is not well defined for IS, as the variance of the raw ratios is infinite when $k>0.5$, although we get finite estimate when running finite number of simulations; see Appendix \ref{sec:truncated_mean}.

\begin{figure}[t]
  \centering
  \includegraphics[height=2.8in]{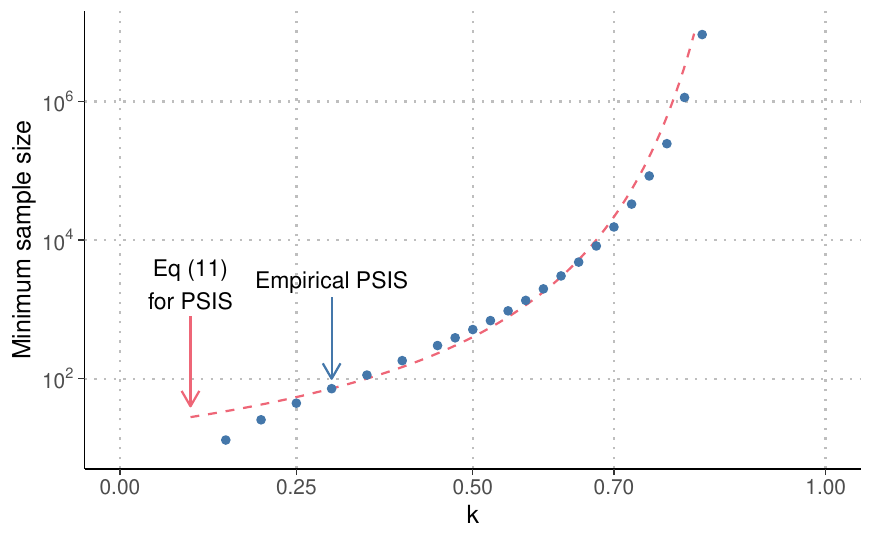}
   \vspace{-.25\baselineskip}
  \caption{{\color{matred}Minimum sample size (red dashed line)} required
    to control RMSE of PSIS as a function of $k$, as computed according to the heuristic
    \eqref{eq:RMSE_minimum_ESS_PSIS}, and {\color{matblue} empirical PSIS result (blue dots)} for the average sample size to obtain fixed RMSE (from 10\,000 repeated simulations). The required sample size quickly grows infeasibly large when  $k>0.7$.}
  \label{fig:minimum_sample_size}
\end{figure}
Appendix \ref{sec:exponential_test} presents more details on the simulation and more results.
Figure \ref{fig:minimum_sample_size} shows the theoretical approximation and empirical result for PSIS RMSE, when the ratios are exactly Pareto distributed.
The simulations match the simple theoretical approximation well even for $k \leq 0.5$, although a different approximation would be more appropriate for $k < 0.5 +0.5/\log_{10}(S)$. 

\subsubsection{Sample Size Dependent $\hat{k}$-threshold}
\label{sec:S-k-threshold}
Based on the bias-variance tradeoff threshold and extensive experiments for moderate sample sizes ($S>2000$), threshold $\hat{k}<0.7$ is a good choice. We can also reverse the minimum sample size requirement to make a sample-size-specific threshold. Based on the experiments, the appropriate constant is such that an easy to remember rule of thumb (for $S \geq 10$) is
\begin{equation}
  \label{eq:max_k}
  1-1/\log_{10}(S).
\end{equation}
Sample sizes 1000, 2000, 4000, and 10\,000 correspond to thresholds $0.67$, $0.7$, $0.72$, and $0.75$, which can all be approximated with a generic threshold $0.7$. For a much smaller sample size 100, the threshold would be $0.5$, and for a much bigger sample size $100\,000$ the threshold would be $0.8$. Although with bigger sample size $S$ we can achieve estimates with small probability of large error, it is difficult to get accurate MCSE estimates as the bias starts to dominate when $k>0.7$ (see Section \ref{sec:PSIS_RMSE}).

\subsubsection{PSIS vs IS L1 Deviation}
\begin{figure}[t]
  \centering
  \includegraphics[height=2.8in]{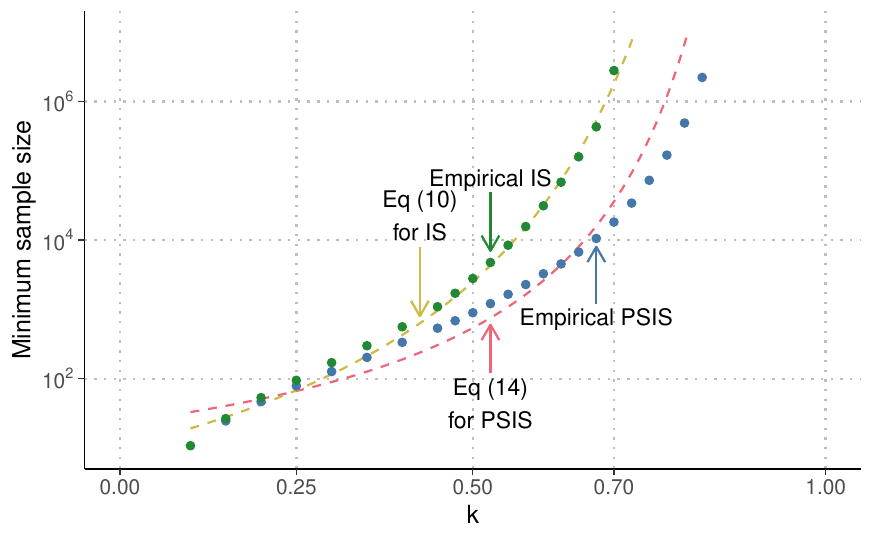}
   \vspace{-.25\baselineskip}
  \caption{{\color{matred}Minimum sample size (red dashed line)} required
    to control L1 deviation by PSIS as a function of $k$, as computed according to the heuristic
    \eqref{eq:L1_minimum_ESS_PSIS}, {\color{matblue} empirical PSIS result (blue dots)} for the average sample size to obtain fixed L1 deviation, {\color{matyellow}minimum sample size (yellow line)} required
    to control L1 deviation by IS as a function of $k$, as computed according to the heuristic
    \eqref{eq:L1_minimum_ESS_IS}, {\color{matgreen} empirical IS result (green dots)} for the average sample size to obtain fixed L1 deviation (from 10\,000 repeated simulations). The required sample size grows more quickly for IS than PSIS, and for PSIS quickly grows infeasibly large when $k>0.7$.}
  \label{fig:L1_theory}
\end{figure}

The normal approximation of the distribution of truncated mean can also be used to approximate quantiles and L1 deviation. \citet{Zaliapin+etal:2005} propose to use this for lower quantiles for which it can be expected to be more accurate, but based on our experiments the approximation is useful to give approximate order of the magnitude for upper quantiles, too. For the raw ratios the quantile was given above as $S^{k-1} (1-q)^{-k}$, but for PSIS we can use approximation
\begin{align}\label{eq:L1_minimum_ESS_PSIS}
S^{k-1} \Phi^{-1}(q).
\end{align}
As normal distribution has shorter tail than a stable distribution with tail index $k>0.5$, we can see that PSIS has smaller L1 deviation than IS, when $k>0.5$.
Figure \ref{fig:L1_theory} shows the empirical and theoretical results for L1 deviation with IS and PSIS ($q=0.99$).

\subsubsection{PSIS RMSE Convergence Rate Given $\hat{k}$}
\label{sec:rate}

In addition to considering how $S$ needs to change to control the error when $k$ increases, it is useful to consider how $S$ needs to change to decrease the error given fixed $k$. PSIS RMSE scaling given different values of $k$  were given in Section \ref{sec:PSIS_RMSE}.

If we consider CLT convergence rate  to have efficiency 1, and write $(S^{-1/2})^\alpha, \alpha=1$, we can calculate the relative efficiency $\alpha$, when $k\in(0.5+0.5/\log_{10}(S), 1)$ and $k=0.5$. By solving $\alpha$ from $S^{k-1}=(S^{-1/2})^\alpha$ and $(S/\log(S))^{-1/2} = (S^{-1/2})^\alpha$, we find that the relative efficiencies for $k\in(0.5+0.5/\log_{10}(S), 1)$ and $k=0.5$ are $2(1-k)$ and $1-1/\log(S)$, respectively (see Appendix \ref{sec:rate_appendix}). The relative convergence rate changes smoothly, but piecewise linear approximation with knots at $0$, $0.5-0.5/\log_{10}(S)$, $0.5$, $0.5+0.5/\log_{10}(S)$, and $1$, is useful for characterizing how the relative convergence rate changes with $k$ (see Appendix \ref{sec:rate_appendix}).

Figure \ref{fig:relative_efficiency} shows theoretical and experimental convergence rate for PSIS RMSE in case of ratio distribution being exactly Pareto distribution. Empirical convergence rate is estimated by dividing the PSIS RMSE of a bigger sample size with the PSIS RMSE of a smaller sample size. The sample sizes in Figure \ref{fig:relative_efficiency} are $3000$ and $10\,000$. Empirical results show that the convergence efficiency starts to drop before $k=0.5$ and matches theoretically predicted efficiency at $k=0.5$. The figure also shows a smooth approximation of the theoretical convergence rate; see the equation and piecewise linear approximation in Appendix \ref{sec:rate_appendix}.
\begin{figure}[t]
  \centering
  \includegraphics[height=2.8in]{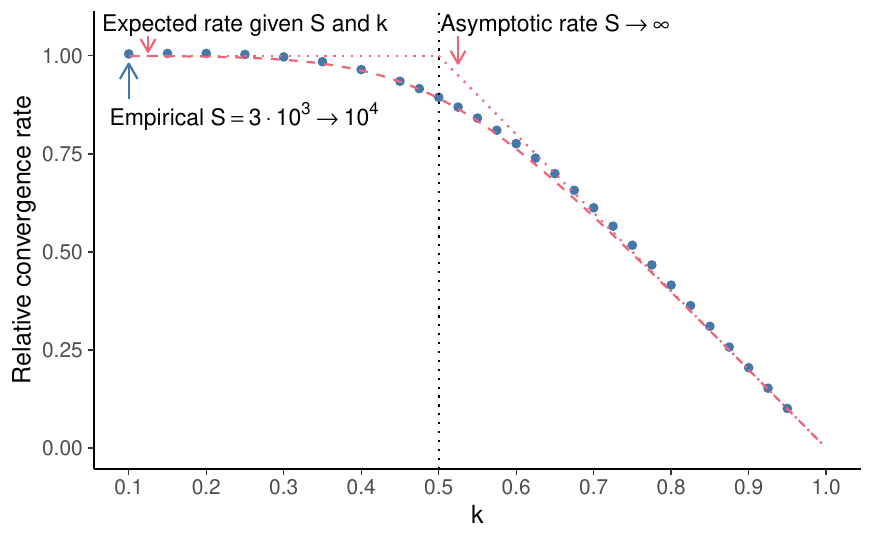}
   \vspace{-.25\baselineskip}
  \caption{Convergence rate as a function of $k$ and $S$. {\color{matred} Red dashed line shows the theoretical convergence rate} based on the CLT and generalized CLT. {\color{matblue} Blue dots show the empirical convergence rate} from the simulation with Pareto distributed ratios (from 10\,000 repeated simulations). Empirical convergence rate is estimated by how much the error decreases when the sample size is doubled $(5\,000 \rightarrow 10\,000)$. For a finite sample size $S$ the transition from CLT convergence rate to GCLT convergence rate is smooth, and the there is no sudden change at $k=0.5$.}
  \label{fig:relative_efficiency}
\end{figure}

Given a sample size $S$, the relative convergence rate results can be used to predict how much bigger sample size is needed to halve the PSIS RMSE. For example if $k=0.7$ approximately $4^{(1/0.7)}\approx 10$ times bigger sample size is needed to halve PSIS RMSE. This empirical result matches the theory well, as the ratio distribution is exactly Pareto distributed. When the ratio distribution is not exactly Pareto distributed, the shape of the bulk and unobserved part of the tail influence the convergence rate, and the above theoretical results should be considered only as approximate guidance. Section \ref{sec:high_dimensions} and Appendix \ref{sec:examples} contain examples of convergence rates when the bulk of the ratio distribution does not follow Pareto distribution.

Appendix B.1 includes more results, and Table \ref{tab:equations_summary} summarizes the useful diagnostic equations discussed above.
\begin{table}
  \centering
  \begin{tabular}{l l l }
    Minimum sample size for reliable Pareto smoothed estimate & $10^{(1/(1-\hat{k}))}$ & (\ref{sec:PSIS_RMSE})\\
    Approximate ESS given $\hat{k}$ & $S/10^{(\hat{k}/(1-\hat{k}))}$ & (\ref{sec:PSIS_RMSE})\\
    Maximum $\hat{k}$ for reliable Pareto smoothed estimate & $1-1/\log_{10}(S)$ & (\ref{sec:PSIS_RMSE}) \\
    Maximum $\hat{k}$ for low bias in Pareto smoothed estimate & 0.7 & (\ref{sec:PSIS_RMSE}) \\
    Convergence rate when $k<0.5-0.5/\log_{10}(S)$ & $\sim S^{1/2}$ & (\ref{sec:rate}) \\ 
    Convergence rate when $k=0.5$ & $S^{(1-1/\log(S))/2}$ & (\ref{sec:rate})\\
    Convergence rate when $k>0.5+0.5/\log_{10}(S)$ & $\sim S^{(1-k)}$ & (\ref{sec:rate}) 
  \end{tabular}
  \vspace{.25\baselineskip}
  \caption{Summary of useful diagnostics related to PSIS RMSE and the convergence rate. The number $10$ used in the four first equations is an ad hoc choice based on rule of $10\%$ and empirical results.}
  \label{tab:equations_summary}
\end{table}

\subsubsection{Only the Tail Has Pareto Shape}
In practice, the distribution of the ratios is not exactly Pareto distributed. If the tail index $k<0.5-0.5/\log_{10}(S)$ (as discussed in Section \ref{sec:PSIS_RMSE}), the variance-based MCSE based on all ratios is accurate, and the exact shape of the ratio distribution has a smaller effect. If the tail index $k$ is close to $0.5$ or larger, the tail behavior dominates the distribution of the mean and the needed sample size can be estimated from $\hat{k}$. The generalized central limit theorem states that mean of any distributions that have Pareto-like tails, converge toward stable distribution with the same tail shape. The result by \citet{Pickands:1975} states that many distributions have tails that can be approximated with the generalized Pareto distribution, which has an additional scale parameter that contributes to the constant $c$ in the needed sample size to obtain a specified error threshold. 

\subsubsection{$h$-specific Estimates}
We smooth only $r(\theta)$, and Section \ref{sec:asymptotic} discusses assumptions needed for $h$, so that PSIS has asymptotic consistency and finite variance. Under these conditions, the above results extend to the analysis of $h(\theta)r(\theta)$, although now there can be a thick tail both left and right, and the tails may have different tail index. The tail with the larger tail index dominates, and thus we set $\hat{k}_h$ to that $\hat{k}$ which is bigger, as discussed in Section \ref{sec:k_hat_big}. In self-normalized importance sampling, there is a ratio of sums. Stability of the ratio will be dominated by the term which has higher $k$ index. For example, if $h(\theta)=\theta$ or $h(\theta)=\theta^2$,
 the numerator has higher $k$ index. In fast leave-one-out cross-validation, $h=1/r$ and the denominator has higher $k$ index.
 The empirical results in this paper support that larger of $\hat{k}_h$ or $\hat{k}$ for the normalization term indicates the overall performance. The empirical results support that despite the deviations from the theory for perfect (truncated) Pareto sums, the diagnostics are sufficiently accurate to be useful.

\subsubsection{Finite Sample Size and Bounded Ratio Distributions}
\label{sec:bounded_ratios}

With finite sample size, the proposal distribution affects how far from the tail of the ratio distribution we are likely to get draws. Previously we used distribution of mean of truncated Pareto variables to explain the behavior of PSIS replacing the largest weights with the expected order statistics. We can use the truncated Pareto to explain also behavior when the raw ratios are bounded, but the bound is much further in tail than the largest observed ratio. 
For a bounded distribution, true $k<0$, all moments are finite, and CLT holds asymptotically. If the bound is far from the observed values, pre-asymptotically the mean behaves statistically as if the distribution has not been truncated \citep[e.g.,][]{Zaliapin+etal:2005}, estimated $\hat{k}$ can be larger than $0$, and high sample size may be needed before the effect of the truncation is observed. A practical consequence is that even if the proposal distribution had been chosen to guarantee bounded ratios, it is possible that we do not observe raw ratios near the bound. We can approximate tails of bounded ratio distributions with a truncated Pareto distribution, and if the bound is larger than any of the observed ratios, we can proceed using $\hat{k}$ diagnostics as discussed above. Furthermore, the ratios of proper densities have finite mean by construction, and thus $k<1$ always, but pre-asymptotically we can observe $\hat{k}>1$, and the sum then behaves statistically as if the distribution has no finite mean. We demonstrate in the next section that in high dimensions, it is likely that the bound is far and the pre-asymptotic diagnostics work well. We illustrate the behavior in case of bounded ratios in Section \ref{sec:high_dimensions} with an example where eventually $\hat{k}>1$, and in Appendix \ref{sec:norm_and_t_test} with an example where the bound is not initially observed, but with increasing sample size is eventually observed.

It would be possible to construct also an artificial example where the
tail would look thin (low Pareto $\hat{k}$) for small $S$ with large
probability, while the true tail is thick (high
Pareto $\hat{k}$). \citet{Chatterjee+Diaconis:2018} present one such
example with a discrete distribution. Except for very small $S$, in all the
simulations and applications where we have applied Pareto $\hat{k}$
diagnostic, we have not observed such behavior.

\subsection{$\hat{k}$ Is a Good Diagnostic in Finite Samples and in High Dimensions}
\label{sec:high_dimensions}

\begin{figure}[tp]
  \begin{center}
    \includegraphics[width=\textwidth]{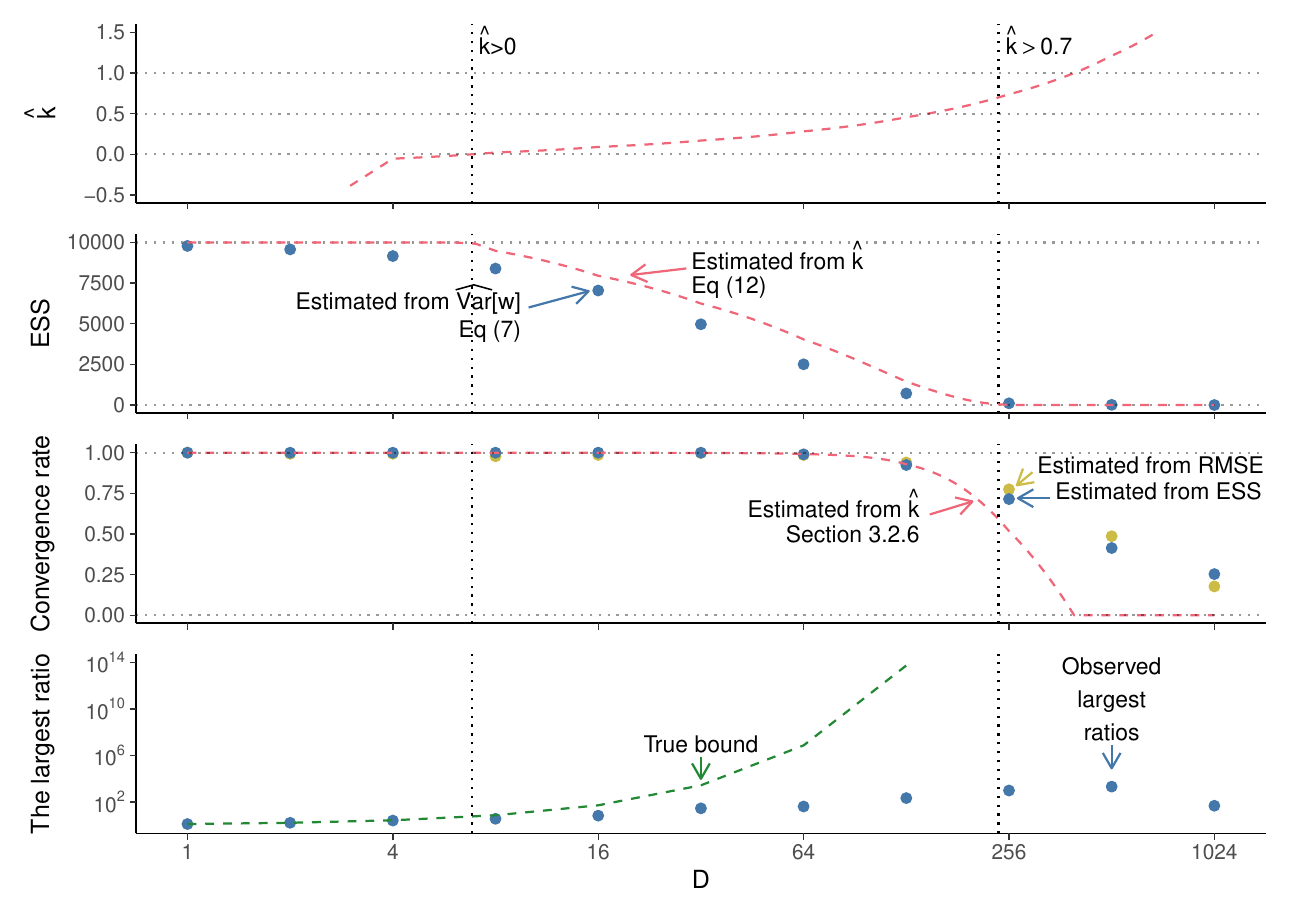}
  \end{center}
  \vspace{-.5\baselineskip}
  \caption{
  {First plot:} {\color{matred}$\hat{k}$ diagnostic}. Asymptotically $k<0$, but with a finite sample size the behavior is statistically as predicted by $\hat{k}$. The vertical dotted lines mark when the tail looks like unbounded $(\hat{k}>0)$, and when the sample size is too small $(\hat{k}>0.7)$.
  {Second plot:} {\color{matblue}estimated effective sample size using Eq.\ \eqref{eq:ESS}} and {\color{matred}estimated effective sample size by $S/10^{(1/(1-k))}$}. By $D=128$, the $S=10\,000$
    importance weighted draws have only a few practically non-zero weights.
  {Third plot:} {\color{matblue}convergence rate estimated by comparing RMSE using $S=5\,000$ and $S=20\,000$} (half and double of original $S$). {\color{matred} Convergence rate estimated from $\hat{k}$}. By $D=512$ the convergence rate is
    about $S^{1/4}$, and getting any improvement in the accuracy becomes tedious. The observed convergence rate is higher than predicted by $\hat{k}$ as the weight distribution is not exactly Pareto shaped and the reduction in bias with increasing sample size dominates. When $\hat{k}>0.7$ the errors are impractically large anyway. The results are similar with L1 deviation and MAE.
    {Fourth plot:} {\color{matgreen}True upper bound of the ratios}. The axis has been truncated at $10^{14}$, and the upper bound for $D=1024$ is approximately $2\cdot 10^{77}$. {\color{matred} The largest observed ratio} given sample of size $S=10\,000$. When the largest observed ratio is far from the bound, $\hat{k}>0$.
  }
  \label{fig:psis_highd_t7v1}
\end{figure}

Although importance sampling papers classically focus on the asymptotic
properties of the estimator, the literature is full of examples where an
importance sampling estimator is simulation-consistent, has finite variance, and is
asymptotically normal but still fails to work. Most of these examples, such as
the one in \citet[][Sec.\ 29.2]{mackay2003information} where the importance
ratios are bounded, occur in high dimensions. Hence, if we want PSIS to work
reliably for problems of any dimension, we need to have a  diagnostic that can
flag poor convergence for any given sample of importance weights.

In the previous section, we argued that if we know the tail behavior of the
importance ratios, we can tell if PSIS will succeed within a reasonable
computational budget. In this section, we argue that the estimate $\hat{k}$ of
the true tail index can quantify the finite sample behavior; see also Section \ref{sec:bounded_ratios}.
It does not matter what the actual tail behavior is if the  distribution of the set of observed importance ratios is heavy tailed.

The following example shows that $\hat{k}$ can be an effective diagnostic for
the real pre-asymptotic convergence behavior. The $\hat{k}$ diagnostic correctly
captures the collapse of the effective sample size \citep[Section
9.3]{Owen:2013} and the convergence rate as the dimension increases.

\begin{example}
Let the target distribution be a $D$-dimensional normal with zero mean and
identity covariance matrix, and let the proposal distribution be a
multivariate Student-$t$ with degrees of freedom $\nu = 7$ and with the same
marginal variance as the target and a diagonal structure matrix.
We take $S=10\,000$ draws from the proposal distribution.
Figure~\ref{fig:psis_highd_t7v1} shows what happens when the number of
dimensions $D$ ranges from $1$ to $1024$. Although the importance ratios are
always bounded and the variance is finite, the finite sample behavior is
indistinguishable from the infinite variance case when the number of dimensions
grows large enough. The effective sample size and convergence rates drop
dramatically, but this can be diagnosed with Pareto $\hat{k}$ diagnostic. In
these cases, we have just not yet reached the asymptotic regime where the
central limit theorem kicks in.
\end{example}

\subsection{Other Minimum Sample Size Estimates}
\label{sec:related_samplesize}

To our best knowledge, we are the first to discuss minimum sample
size estimates for importance sampling with truncated weights as in PSIS and TIS. Here we discuss Kullback-Leibler divergence and max-sum based results that are applicable for IS with non-truncated raw weights.

Several authors
\citep{Sanz-Alonso:2018,agapiou2017importance,Chatterjee+Diaconis:2018},
have proved, under some assumptions, that the necessary and sufficient sample size to
control error of IS with raw weights is roughly
\begin{equation}\label{eq:expKL}
  \exp\left(\KL(p||g)\right)
  = \exp\left( \int p(\theta)\log\left(\frac{p(\theta)}{g(\theta)}\right) d\theta\right),
\end{equation}
where $\KL(p||g)$ is the Kullback-Leibler divergence from $p$ to $g$.
The most comprehensive results are due to \citet{Chatterjee+Diaconis:2018}, who show
that a larger sample size than this gives tail guarantees for the error in
self-normalized importance sampling, while for smaller sample sizes it is not
possible to control the large deviations.

In general, we are not able to compute $\KL(p||g)$ in \eqref{eq:expKL} directly. However, we can turn \eqref{eq:expKL} into a heuristic bound. 
Assume that the density ratios $r(\theta) =  p(\theta)/g(\theta)$ exactly form a
generalized Pareto distribution \eqref{pareto} with the location parameter $u=0$, scale parameter
$\sigma>0$, and shape parameter $k\in (0,1)$, when $\theta$ is distributed by $g$.  From here,  we compute
$$
\KL(p||g) = \frac{\sigma}{1-k}\left(1-\gamma- {\psi}\left(-1+\frac{1}{k}\right)  - \log\left(\frac{k}{\sigma}\right) \right),
$$
where $\gamma \approx 0.577$ is the Euler constant and
$\psi(x)= \frac{d}{dx} \log \Gamma(x)$ is the digamma
function. Expanding the digamma function near 0, we obtain
$\KL(p||g) =  (1-k)^{-1}+ \mathcal{O}((1-k)^{-2})$ as
$k \to 1^-$. Thereby the necessary and sufficient sample size has the limiting order 
$$\log S\approx (1-k)^{-1}$$ when $k \to 1^-$, matching the limiting order of $\log S$ in Eqs.\ \eqref{eq:L1_minimum_ESS_IS} and \eqref{eq:RMSE_minimum_ESS_PSIS}.

Noting that the sample variance of IS raw weights can be a bad measure of its performance, \cite{Chatterjee+Diaconis:2018} proposed an alternative measure of quality,
\begin{align*}
Q_S =\frac{ \max_{1\leq s \leq S}     r(\theta_s)       }{\sum_{s=1}^S      r(\theta_s)  },
\end{align*}
where the sample size is considered large enough if $Q_S$ is below some pre-specified threshold (e.g., 0.01).  Heuristically, when sample size is not large enough,  the largest weight dominates the summation, making $Q_S$ close to 1. \citet{Zaliapin+etal:2005} examine $Q_S^{-1}$ for Pareto distributed variables, and provide a large-sample approximation for its expectation (Eq.\ 34)
\begin{align*}
\E\left( Q_S^{-1} \right) \approx S^{1-k}\Gamma(k+1)/(1-k),   \quad 0<k<1.
\end{align*}
Inverse of this has the same order as Eq.\ \eqref{eq:Pareto_mean_quantile}, and the derived estimate for the minimal sample size required to reach the pre-specified threshold has the same order as Eq.\ \eqref{eq:L1_minimum_ESS_IS}. \citet{Zaliapin+etal:2005} discuss and demonstrate that in practice $Q_S^{-1}$ has high variation even with large $S$, which is problematic when we assume that only the tail can be approximated with Pareto distribution.

\section{Convergence of PSIS}
\label{sec:asymptotic}

In this section, we present the result that PSIS is asymptotically simulation-consistent and has finite
variance under some standard conditions. In actual PSIS,
$k$ and $\sigma$ are estimated consistently when $M\rightarrow\infty$ and $M/S \rightarrow 0$, but
we focus on an idealized variant of PSIS where the 
parameters $k$ and $\sigma$ are fixed, although they may be different to the true tail parameters.
For simplicity, we limit ourselves to ordinary PSIS, although consistency of self-normalized PSIS follows from Slutsky's theorem
by the same arguments as in TIS \citep[][Appendix B]{Ionides:2008}.

The theory in this section focuses on asymptotic properties, while
Section \ref{sec:k_hat_big} discusses pre-asymptotic behavior, and we demonstrated in Section \ref{sec:high_dimensions} that reaching the  asymptotic regime can require infeasible
sample size even in
a simple example.

For the purpose of this section, we will take $M = \sqrt{S}$, which we will
assume to be an integer. Asymptotically, there is no difference between this
and taking $M=\lfloor 3\sqrt{S}\rfloor$, but the notation is simpler.

\subsection{Asymptotic Consistency}

Intuitively we get asymptotic consistency as $M/S \rightarrow 0$ when $S \rightarrow \infty$, and the effect of smoothing eventually vanishes, and thus the usual consistency of importance sampling holds.

For more detailed conditions, 
consider the order statistics $r_{1:S} \leq r_{2:S}\leq \dots \leq r_{S:S}$ of
$r(\theta_s)$. For convenience we reorder the sample so that
$r(\theta_s) = r_{s:S}$.  It is also convenient to rewrite the weights on the
upper tail as
$$
w_j = F^{-1}\left(\frac{j-1/2}{M}\right)
  = r_{(S-M+1:S)} +  \frac{\sigma}{k}\left(\left(1-\frac{j-1/2}{M}\right)^{-k} - 1\right)
  = r_{(S-M+1:S)} + w^{+}_j,
$$
where the $w^{+}_j$'s are deterministic (given $k$ and $\sigma$) and $j=1,\dots,M$.

We can then write the PSIS estimator as a Winsorized version of the truncated
estimator of \citet{Ionides:2008} plus an extra bias correction term:
$$
I^S_h =
  \frac{1}{S} \sum_{s=1}^{S}\left(r(\theta_s) \wedge r_{(S-M+1):S}\right)h(\theta_s)
  \,+\, \frac{1}{S}\sum_{j=1}^M w^{+}_j h(\theta_{S-M+j }).
$$
The following theorem, which is proved in  Appendix \ref{appendix:proof}, shows that our idealized version of PSIS converges under mild conditions.
\begin{theorem} \label{thm:convergence}
Let $\theta_s$, $s = 1,\ldots, S$ be an iid sample from $G$ and let $r_s = r(\theta_s) \sim R$. Assume that
\begin{enumerate}
\item  $R$ is absolutely continuous,
\item $M  = \mathcal{O}(S^{1/2})$,
\item $h \in L^2(p)$, and
\item $k$ and $\sigma$ are known, with $\sigma = \mathcal{O}(r_{S-M+1:S})$.
\end{enumerate}
Then Pareto smoothed importance sampling converges in $L^1$ and its variance goes to zero. It is, therefore, consistent and asymptotically unbiased.
\end{theorem}

\subsection{Asymptotic Normality}

Considering PSIS bias correction as the mean of truncated Pareto
distributed variables (Section \ref{sec:k_hat_big}), we can assume
asymptotic normality, but as the truncation of that Pareto
distribution depends on the ratios themselves, more detailed analysis
would be required  to establish the conditions.

\citet{griffin1988asymptotic} suggests that if the product
$(h(\theta_s)r(\theta_s))$ is Winsorized at both ends, the von Mises
condition imply that the Winsorized estimator is asymptotically
normal. It seems likely that this can be shown also for PSIS,
but that is an open question.

\section{Practical Examples}\label{sec:practical}

In this section, we present three examples where Pareto smoothed importance
sampling improves the estimates and where the Pareto shape estimate $\hat{k}$ is
a useful diagnostic. In the first example PSIS is used to improve the
distributional approximation (split-normal) of the posterior of a logistic
Gaussian process density estimation model. We then demonstrate the performance
and reliability of PSIS for leave-one-out cross-validation (LOO-CV) analysis of
Bayesian predictive models for the canonical stacks data as well as for a recent
breast cancer tumor data set with 105 different protein expressions. Further
examples with simulated data can be found in Appendix \ref{sec:examples}.

\subsection{Improving Distributional Approximation with Importance Sampling}
\label{sec:distributional_example}

The first example shows that PSIS can be useful for performing approximate
Bayesian inference. PSIS has been used to improve and diagnose variational
approximations to posteriors
\citep{Yao+Vehtari+Simpson+Gelman:2018,Magnusson+etal:2019:adviloo,Magnusson+etal:2020,Dhaka+etal:2020,Dhaka+etal:2021:challenges-in-vi,Zhang+etal:2022:pathfinder}.  In this
section, we show that PSIS can be used to speed up logistic Gaussian process
(LGP) density estimation \citep{Riihimaki+Vehtari:2014}, which is implemented in
the   GPstuff toolbox\footnote{Code available at
\url{https://github.com/gpstuff-dev/gpstuff}}\citep{Vanhatalo+gpstuff:2013}.

LGP provides a flexible way to define the smoothness properties of density
estimates via the prior covariance structure, but the computation is
analytically intractable. \citet{Riihimaki+Vehtari:2014} propose a fast
computation using discretization of the normalization term and Laplace's method
for integration over the latent values.

Given $n$ independently drawn $d$-dimensional data points
$x_1,\ldots,x_n$ from an unknown distribution in a finite region
(having a compact support) $\mathcal{V} \subset \Re^d$, we want to
estimate the density $p(x)$.
To introduce the constraints that the density is non-negative and that its
integral over $\mathcal{V}$ is equal to 1, \citet{Riihimaki+Vehtari:2014}
employ the logistic density transform, 
\begin{equation}\label{eq_logdens}
p(x)=\frac{\exp(f(x))}{\int_{\mathcal{V}}\exp(f({s}))d{s}},
\end{equation}
where $f$ is an unconstrained latent function.  To smooth the density estimates,
a Gaussian process prior is set for $f$, which allows for assumptions about the
smoothness properties of the unknown density $p$ to be expressed via the
covariance structure of the GP prior. To make the computations feasible,
$\mathcal{V}$ is discretized into finite $m$ subregions (or intervals if the
problem is one-dimensional).
Here we skip the details of the Laplace approximation and focus on the importance
sampling.

Following \citet{Geweke:1989}, \citet{Riihimaki+Vehtari:2014} use importance
sampling with a multivariate split Gaussian density as an approximation. The
approximation is based on the posterior mode and covariance, with the density
adaptively scaled along principal component axes, in positive and negative
directions separately, to better match the skewness of the target distribution; see also
\citet{Villani+Larsson:2006}. To further improve the performance,
\citet{Riihimaki+Vehtari:2014} replace the discontinuous split Gaussian used by
Geweke with a continuous version.

\citet{Riihimaki+Vehtari:2014} use an ad hoc soft thresholding of the
importance weights if the estimated effective sample size as defined
by \citet{Kong+Liu+Wong:1994} is less than a specified threshold. The
approach can be considered to be a soft version of truncated importance sampling,
which \citet{Ionides:2008} also mentions as a possibility without
further analysis.
Here we propose to use PSIS to stabilize the weights.

We repeat the density estimation using the Galaxy data set\footnote{\url{https://stat.ethz.ch/R-manual/R-devel/library/MASS/html/galaxies.html}}
1000 times with different random seeds. The model has 400 latent values, that
is, the posterior is 400-dimensional, although due to a strong dependency
imposed by the Gaussian process prior the effective dimensionality is smaller.
Because of this, it is sufficient that the split-normal is scaled only along the
first 50 principal component axes.
In order to compare to a baseline method, we implement the Markov chain Monte
Carlo scheme described in \citet{Riihimaki+Vehtari:2014}.
Computation time for MCMC inference was about half an hour and
computation time for split-normal with importance sampling was about
1.3 s (laptop with Intel Core i5-4300U CPU @ 1.90GHz x 4).

\begin{figure}[t]
  \begin{center}
    \includegraphics[]{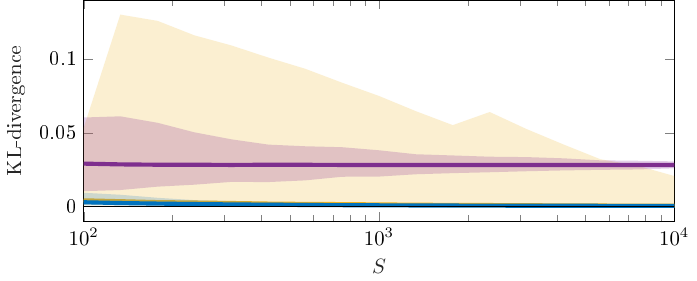}
  \end{center}
   \vspace{-.5\baselineskip}
  \caption{Kullback-Leibler divergence from the density estimate using MCMC
  to the density estimates using {\color{matpurple} the plain split-normal approximation (purple)},
  {\color{matyellow} IS (yellow)}, and {\color{matblue} PSIS (blue)}. The shaded areas show the envelope of the
  KL-divergence from all 1000 runs. The variability of {\color{matpurple} the plain split-normal
  approximation (purple)} reduces with increasing number of draws $S$, but the
  KL-divergence doesn't decrease. {\color{matyellow} IS (yellow)} has the highest variability.
  }
   \label{fig:lgpdens_kl_conver}
\end{figure}

Figure \ref{fig:lgpdens_kl_conver} compares the Kullback-Leibler divergence from
the density estimate using MCMC to the density estimates using the split-normal
approximation with and without the importance sampling correction.
The shaded areas show the envelope of the KL-divergence from all 1000 runs. The
variability of the plain split-normal approximation (purple) diminishes as the
number of draws $S$ increases, but the KL-divergence does not decrease. IS
(yellow) has high variability. PSIS (blue) performs well, with a small
KL-divergence already when $S$ is only $100$. TIS results (not shown) were
mostly similar to PSIS, with some occasional worse results (similar jumps as in
Figure~\ref{fig:dimxoffnt20s_error0}). The mean estimate for $\hat{k}$ was
$0.43$ with $S=100$ and $0.55$ with $S=10^4$, which explains the high
variability of IS, the occasional bad results from TIS, and the excellent performance
of PSIS. These $\hat{k}$ values also signal that we can trust the PSIS results.

The Pareto $\hat{k}$ diagnostic can also be used to compare the quality of the
distributional approximations. In the case of a simple normal approximation
without split scaling, the mean $\hat{k}$ with $S=10^4$ was $0.60$, and thus
slightly higher variability and slower convergence can be assumed relative to
the split-normal approximation.

\subsection{Importance-sampling Leave-one-out Cross-validation}
\label{sec:PSIS-LOO}

We next demonstrate the use of Pareto smoothed importance sampling for
leave-one-out cross-validation (LOO) approximation.
The $i$th leave-one-out cross-validation predictive density can be approximated with
\begin{align}
  \label{eq:is-loo}
p(\tilde{y}_i|y_{-i}) \approx
\frac{\sum_{s=1}^S w_{i}(\theta_s) p(\tilde{y}_i|\theta_s)}{\sum_{s=1}^S w_{i}(\theta_s)}.
\end{align}
Importance sampling LOO was proposed by \citet{Gelfand+Dey+Chang:1992}, but for long time it
was not widely used as the estimator is unreliable if the weights
have infinite variance. For some simple models, such as linear and
generalized linear models with specific priors, it is possible to
analytically check the 
sufficient conditions for the
variance of the importance weights in IS-LOO to be finite
\citep{Peruggia:1997,Epifani+MacEachern+Peruggia:2008}, but this is
not generally possible. Furthermore, even if the variance would be finite, it is possible that the pre-asymptotic behavior is indistinguishable from the infinite variance case as discussed in Section \ref{sec:k_hat_big}.

We first demonstrate properties of IS, TIS, and PSIS with the stack loss data,
which is known to have one observation producing infinite
variance for LOO importance ratios. Then we demonstrate the speed and
reliability of PSIS-LOO for performing model assessment and comparison
for predictive regression models for 105 different protein expressions.

\subsubsection{LOO for Stack Loss Data}
The stack loss data has $n = 21$ daily observations on one response
variable and three predictors pertaining to a plant for the oxidation
of ammonia to nitric acid.
The model is a simple Gaussian linear regression. We fit the model using
Stan \citep{Stan:2017}; the code is in appendix \ref{sec:stan_stack_code}.
\citet{Peruggia:1997} showed,  for a specific choice of prior distributions that we do not
recommend to be used in real analyses, that the importance ratios have an
infinite variance when leaving out the first data point.

\begin{figure}[t]
  \begin{center}
    \includegraphics[width=\textwidth]{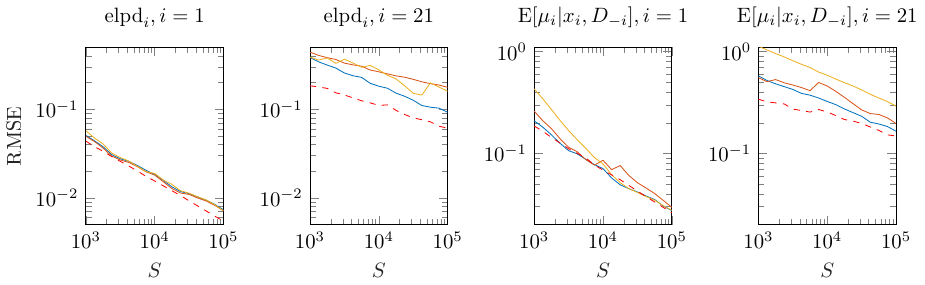}
  \end{center}
   \vspace{-0.5\baselineskip}
  \caption{RMSE with {\color{matyellow}IS (yellow)}, {\color{matred}TIS (red)} and {\color{matblue}PSIS (blue)}, and {\color{red}the MCSE
  estimates with PSIS (red dashed lines)} for the expected log predictive
  densities $\mathrm{elpd}_i=\log p(y_i|x_i,D_{-i})$ and leave-one-out
  predictive mean $\E(\mu_i|x_i,D_{-i})$. Average $h$ specific $\hat{k}$'s,
  using $S=10^5$ draws from each of 100 runs, are $0.46, 0.79, 0.45, 0.81$,
  in the order of subplots.}
   \label{fig:stacks_loo_errorsb}
\end{figure}
Figure~\ref{fig:stacks_loo_errorsb} shows the RMSE (compared to the exact LOO) and MCSE estimate (combined error from importance sampling and MCMC as defined in Section \ref{s:pareto}) from
100 runs for the LOO estimated expected log predictive densities
$\mathrm{elpd}_i=\log p(y_i|x_i,D_{-i})$ and leave-one-out predictive mean
$\E(\mu_i|x_i,D_{-i})$ (where $D_{-i}=(x_{-i},y_{-i})$ denotes the data without $i$th
observation) estimated with IS, TIS and PSIS when leaving out the first or 21st
observation. Pareto smoothing and MCSE estimates were adjusted
based on relative MCMC sample efficiency. The true values were computed by
actually leaving out the $i$th observation and using multiple long MCMC chains
to get a small Monte Carlo error.  PSIS gives the smallest RMSE, and
the accuracy of the MCSE estimates are what we would expect based
on $h$ specific $\hat{k}$'s:  error estimates are accurate for
$\hat{k}<0.5$ and optimistic for $\hat{k}>0.7$.

\subsubsection{LOO for 105 Protein Expression Data Sets}

We demonstrate the benefit of fast importance sampling leave-one-out
cross-validation and PSIS diagnostics with the example of a model for the
combined effect of microRNA and mRNA expression on protein expression. The data
were published by \citet{Aure2015} and are publicly available; we used the
preprocessed data as described by \citet{aittomaki2016thesis}. Protein, mRNA,
and microRNA expression were measured from 283 breast cancer tumor samples, and
when predicting the protein expression the corresponding gene expression and 410
microRNA expressions were used.
We assumed a multivariate linear model for the effects with a Gaussian prior and used Stan
\citep{Stan:2017} to fit the model.
Initial analyses gave reason to suspect outlier observations; to
verify this, we compared Gaussian and Student-$t$ observation models.

For 4000 posterior draws, the computation for one gene and one model took about
9 minutes (desktop Intel Xeon CPU E3-1231 v3 @ 3.40GHz x 8), which is reasonable
speed. For all 105 genes, the computation took about 30 hours. Exact
regular LOO for all models would have taken 125 days, and 10-fold cross-validation for
all models would have taken about 5 days. Pareto smoothed importance sampling LOO
(PSIS-LOO) took less than one minute for all models. However, we do get several
leave-one-out cases where $\hat{k}>0.7$, which we should not trust based on our
results above.
\begin{figure}[tp]
  \begin{center}
    \includegraphics[]{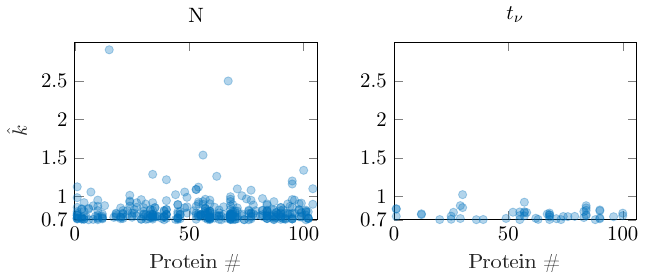}
  \end{center}
   \vspace{-0.75\baselineskip}
  \caption{{\color{matblue} $\hat{k}>0.7$} values for 105 Gaussian and Student-$t$ linear
  models predicting protein expression levels.}
   \label{fig:varsel_khats}
\end{figure}
Figure~\ref{fig:varsel_khats} shows $\hat{k}>0.7$ values for 105 Gaussian and
Student-$t$ linear models, where each model may have several leave-one-out cases
with $\hat{k}>0.7$. Large $\hat{k}$ values arise when the proposal and the
target distributions are very different, which is typical when there are highly
influential observations. Switching to the Student-$t$ model reduces the number
of high $\hat{k}$ values, as the outliers are less influential if they are far
in the tail of the $t$-distribution. When working with many different models, the
Stan team has noticed that high $\hat{k}$ values are also a useful indicator
that there is something wrong with the data or the model \citep{gabry2019}.

\begin{figure}[tp]
  \begin{center}
    \includegraphics[]{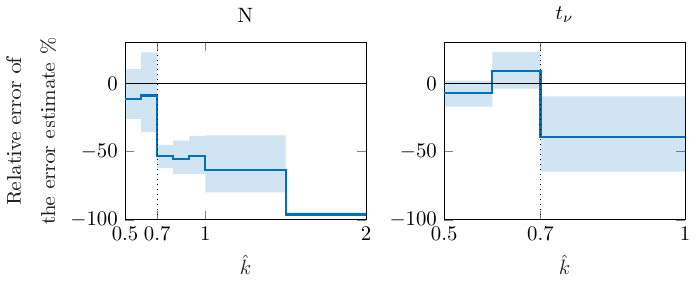}
  \end{center}
   \vspace{-0.75\baselineskip}
  \caption{{\color{matblue} Relative error} of the PSIS MCSE estimates. True
  error was computed as RMSE of expected log predictive density
  $\mathrm{elpd}_i$ estimates with $\hat{k}$ on common interval.  Boundaries of
  intervals can be seen as steps in the plots.}
   \label{fig:varsel_relerror}
\end{figure}
Figure \ref{fig:varsel_relerror} shows the accuracy of PSIS MCSE
estimates for the expected log predictive densities with respect to different
$\hat{k}$ values (computed only for $\hat{k}>0.5$). True values were computed by
actually leaving out the $i$th observation and rerunning MCMC.  We can see that
$\hat{k}$ is a useful diagnostic and the MCSE estimates are
accurate for $\hat{k}<0.7$, as in the simulation experiments, and fail for
$\hat{k} \geq 0.7$.

To improve upon PSIS-LOO we can make the exact LOO computations for any points
corresponding to $\hat{k}>0.7$ (for which we cannot trust the MCSE
estimates). In this example, there were 352 such cases for the Gaussian models
and 53 for the Student-$t$ models, and the computation for these took 42 hours.
Although combining PSIS-LOO with exact LOO for certain points substantially
increases the computation time in this example, it is still less than the time
required for 10-fold-CV.

\begin{figure}[tp]
  \begin{center}
    \includegraphics[width=0.95\textwidth]{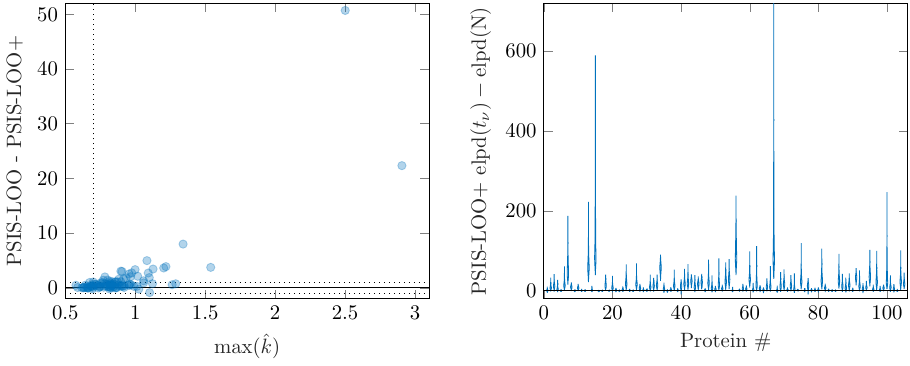}
  \end{center}
   \vspace{-.5\baselineskip}
  \caption{Left plot shows comparison of PSIS-LOO and PSIS-LOO+
  (PSIS-LOO with exact computation for cases with $\hat{k}>0.7$) when comparing
  the difference of expected log predictive densities
  $\sum_{i=1}^n(\mathrm{elpd_i(t_\nu)}-\mathrm{elpd_i(\N)})$. Right plot
  shows the PSIS-LOO+ estimated improvement of expected log predictive densities
  when switching from Gaussian model to Student-$t$ model.}
   \label{fig:varsel_pluscomparison_violin}
\end{figure}
The left subplot in Figure~\ref{fig:varsel_pluscomparison_violin} shows
comparison of PSIS-LOO and PSIS-LOO+ (PSIS-LOO with exact computation for cases
with $\hat{k}>0.7$) when comparing the difference of expected log predictive
densities $\sum_{i=1}^n\left(\mathrm{elpd}_i(t_\nu)-\mathrm{elpd}_i(\N)\right)$.
We see that with high $\hat{k}$ values, the error of PSIS-LOO can be large (the
error would be large for IS and TIS, too). To trust the model comparison, we
recommended using PSIS-LOO+ with exact computation for cases with $\hat{k}>0.7$
or $K$-fold-CV. The right subplot shows the final model comparison results for
all 105 models predicting protein expression levels. For most of the proteins,
the student-$t$ model is much better, and the Gaussian model is not
significantly better for any of the proteins.

\section{Discussion}
\label{s:discussion}

Importance weighting is a widely used tool in statistical computation. Even in
the modern era of Markov chain Monte Carlo, approximate algorithms are often
necessary, in which case we should adjust approximations when possible to better
match target distributions. However, a challenge for practical applications of
importance weighting is the well known fact that importance-weighted estimates
are unstable if the weights have high variance.

In this paper, we have shown that it is possible to reduce the mean square error
of importance sampling estimates using a particular stabilizing transformation
that we call Pareto smoothed importance sampling (PSIS). The key step is to
replace the largest weights by expected quantiles from a generalized Pareto
distribution. We have also demonstrated greatly improved Monte Carlo standard error
estimates, natural diagnostics for gauging the reliability of the estimates, and
empirical convergence rate results that closely follow known theoretical
results.

Pareto $\hat{k}$ diagnostic allows the
user to assess the reliability of PSIS:
\begin{itemize}
\item If $\hat{k} < 1 - 1/\log_{10}(S)$, the PSIS estimate is expected
  to be accurate;
\item If $\hat{k} < \min(1 - 1/\log_{10}(S), 0.7)$, PSIS MCSE estimate
  is expected to be reliable;
\item If $\hat{k} > 0.7$, it quickly becomes too expensive to get an
  accurate estimate;
\item If $\hat{k} > 1$, it is expected that the mean does not exist and
  any estimate for the mean is invalid.
\end{itemize}
With these recommendations, PSIS is a reliable, accurate, and trustworthy
variant of importance sampling that comes with a built-in heuristic that allows
it to fail loudly when it becomes unreliable.

Although $\hat{k}$ was originally targeted for self-normalized importance sampling, it can be used to diagnose also ordinary importance sampling and the distribution of any Monte Carlo draws as discussed and demonstrated by some references mentioned in the next section. 

\subsection{Other Examples in the Literature}

Since the appearance of the original preprint in 2015
\citep{psis:2015}, this work has been cited over 300 times, and
during that time there have been a number of extensions and
applications proposed in the literature.  In this section, we
highlight the main ones we are aware of.

First, there is a suite of works that investigate the use of PSIS 
for computing cross-validation scores. 
\citet{Vehtari+Gelman+Gabry:2017} provide additional comparisons
between IS-LOO, TIS-LOO, PSIS-LOO, and the widely applicable information
criterion \citep{Watanabe:2010d}. The PSIS-LOO method has been implemented, for example, in the
\textsc{loo} package \citep{loo-software:2024}, \textsc{ArviZ}
\citep{Kumar+etal:2019:ArviZ}, and \textsc{Pyro} \citep{bingham2019pyro}.
\citet{Vehtari+etal:2016:loo-glvm} use PSIS-LOO to improve integrated
importance sampling in case of Gaussian process models with
non-Gaussian observation models (in the \textsc{GPstuff} package).
\citet{Buerkner+Gabry+Vehtari:2020:loo-non-factorizable} use PSIS-LOO for Bayesian
non-factorized multivariate normal and Student-$t$ models (in the \textsc{loo} package).
\citet{Magnusson+etal:2019:adviloo,Magnusson+etal:2020} apply PSIS-LOO
in case of variational and Laplace approximations to posteriors (in the \textsc{loo} package).
\citet{Buerkner+etal:2020:LFO} apply PSIS in leave-future-out
cross-validation for time series, where the proposal distribution is
conditioned on fewer data and thus by construction tend to have thicker
tails than the target distribution, but with a large number of steps
in time, eventually new MCMC posterior computation is needed (in \textsc{loo} package).
\citet{Fong+Holmes:2021:conformal} comment how PSIS could improve
add-one-in importance sampling for conformal Bayesian
computation. In add-one-in
importance-sampling one of the existing observations is included again
to update the posterior.
\citet{Silva+Zanella:2023:MixIS} use the Pareto-$\hat{k}$ diagnostic to
compare use of full data posterior and mixture of leave-one-out
posteriors as the proposal in PSIS-LOO.
\citet{Chang+etal:2024:gradient_PSISLOO} use the Pareto-$\hat{k}$
diagnostic to compare use of full data posterior and gradient-flow
based transformed posterior as the proposal in PSIS-LOO.

Second, there has been work using PSIS as a building block to improve
Bayesian models and workflow.
\citet{Yao+etal:2018:stacking,Yao+etal:2021:hierstacking,Yao+etal:2020:multistacking}
use PSIS-LOO for Bayesian stacking (in \textsc{loo} package), which is a form of model combination common in 
machine learning, and for its extensions to hierarchical stacking
and stacking of non-mixing MCMC chains.
\citet{Piironen+Paasiniemi+Vehtari:2020:projpred,projpred,McLatchie+etal:2023:projpred_workflow} use PSIS-LOO as part
of projection predictive variable selection to speed-up the selection of
the number of variables to include (in \textsc{projpred} package).
\citet{McCartan:2021:adjustr} uses PSIS for prior sensitivity
analysis (in the \textsc{adjustr} package), and \citet{Kallioinen+etal:2024:priorsense} use PSIS and
iterative moment matching \citep{Paananen+etal:2021:implicit} for diagnosing both prior and
likelihood sensitivity (in the \textsc{priorsense} package).
\citet{Timonen+etal:2023:PSIS-ODE} use PSIS to allow use of faster ordinal differential equation solving within Bayesian models, which can speed up posterior inference a few orders of magnitude, while providing diagnostic to assess the reliability of the inference (in the \textsc{odemodeling} package).

Third, the $\hat k$ diagnostic has been used to compare variational
posterior approximations, for example, by
\citet[][in \textsc{Stan} software]{Yao+Vehtari+Simpson+Gelman:2018},
\citet{Dhaka+etal:2021:challenges-in-vi},
\citet[][in \textsc{Stan} software]{Zhang+etal:2022:pathfinder}, and
\citet{Liang+etal:2023:heavy_tailed}.

Finally, $\hat{k}$ can be used to diagnose any Monte Carlo estimate.
\citet{Dhaka+etal:2020} use $\hat{k}$ to diagnose number of finite
fractional moments of stochastic optimization in approximate
convergence.
\citet{Paananen+etal:2021:implicit} discuss and demonstrate that
Pareto $\hat{k}$ is useful for any Monte Carlo expectation estimate to
diagnose whether the sampled distribution has tail with $\hat{k}<0.7$.
\citet{Dhaka+etal:2021:challenges-in-vi} use $\hat{k}$ to diagnose the
behavior of Monte Carlo estimates of commonly used divergences in
variational inference.
All the diagnostics described in Section \ref{sec:k_hat_big} have been
implemented, for example, in the \textsc{posterior} package
\citep{posterior-software:2023}.

\subsection{Other Weight Transformations}

Truncated weights have been studied also by
\citet{Koblents+Miguez:2015:PMC_transformed_weights}, who use
condition $M/S \rightarrow 0$ in the asymptotic proofs, but in
practice recommend to set $M=S/10$. This approach is likely to have
similar bias as TIS by \citet{Ionides:2008}, as the difference is just
in the rule to choose the truncation point.

\citet{Miguez:2017:PMC_performance_transformed_weights} propose to set
the largest weights above some quantile to the average of those
weights. This reduces the bias compared to the simple truncation. This
approach keeps the average of all weights the same, so it has a
smaller bias than TIS (given a similar truncation rule). Compared to
PSIS, this approach is likely to (a) have higher variance as the
highest weight can still have a big influence on the average of the
largest weights, (b) have higher bias as on expectation the smallest of the
transformed weights are overweighted and largest of the transformed
weights are underweighted, and (c) be more sensitive to the threshold
as the largest non-transformed weight and smallest transformed weight
are made more different.

\citet{Martino+etal:2018:weight_clipping_comparison} compare the
methods by \citet{Koblents+Miguez:2015:PMC_transformed_weights} and
\citet{Miguez:2017:PMC_performance_transformed_weights}. They state
that condition $M \leq \sqrt{S}$ is useful for the asymptotic results,
but in the experiments set $M=S/5$.

\subsection{Multiple and Adaptive Importance Sampling}
Pareto $\hat{k}$ can be used to compare proposal distributions
both in single and multiple importance sampling \citep[for a review of
multiple IS, see][]{Elvira+etal:2019:generalized_mis}, and as a
diagnostic in adaptive importance sampling \citep[for a review
of adaptive IS, see][]{Bugallo+etal:2017:adaptive_is}.
Sometimes it can be possible to choose a proposal distribution that
guarantees finite variance of importance weights \citep[see,
e.g.,][]{Owen+Zhou:2000:safe_is}, but this doesn't guarantee useful
pre-asymptotic convergence rate as shown in
Section~\ref{sec:high_dimensions}.

In Section~\ref{sec:distributional_example} we demonstrated use of
$\hat{k}$ to compare simple normal 
and split-normal proposal
distributions. \citet{Dhaka+etal:2021:challenges-in-vi} use $\hat{k}$
to compare normal, Student-$t$, planar flow
\citep{Rezende+Mohamed:2015:planar-flow} and NVP flow
\citep{Dinh+etal:2017:nvp-flow} variational approximations in low and high dimensions.

\citet{Paananen+etal:2021:implicit} use $\hat{k}$ as part
of the implicitly adaptive multiple importance sampling approach to decide
when the iterative adaptation can be stopped and thus minimizing the
computational cost (in the \textsc{loo} and \textsc{iwmm} packages).
\citet{Yao+etal:2020:adaptive_path_sampling} use $\hat{k}$
as part of the adaptive path sampling algorithm to decide whether further
adaptation of the pseudo-prior is required.

In all these examples, PSIS was also used to smooth the weights and
improve the final importance sampling estimates.

An anonymous reviewer suggested that $\hat{k}$ could be used
also as a diagnostic tool to decide which weighting schemes
\citep[see][]{Elvira+etal:2019:generalized_mis} are better, or to
select the operation point in the tradeoff between variance reduction
in the weights and computational complexity; see
\citet{Elvira+etal:2015:efficient_mis}.

\subsection{Population Monte Carlo and Particle Filters}

PSIS is especially useful when one proposal distribution or its
computationally cheap transformation can be used for several slightly
different target distributions, as in leave-one-out \citep{Vehtari+Gelman+Gabry:2017} and
leave-future-out cross-validation \citep{Buerkner+etal:2020:LFO}.
In leave-future-out
cross-validation, the observations are added sequentially, and the
posterior approximation is updated using importance weighting until
$\hat{k}>0.7$ and then the posterior sample is regenerated with
MCMC. This resembles the algorithms such as population and sequential Monte Carlo, and particle filtering \citep[e.g.,][part
8]{Cappe+etal:2004:PMC,Crisan+Rozovskii:2011:nonlinear-filtering}.

\citet{Koblents+Miguez:2015:PMC_transformed_weights} and
\citet{Miguez:2017:PMC_performance_transformed_weights} demonstrate
the benefit of variants of truncated importance sampling in population
Monte Carlo, and it can be assumed that PSIS would further improve the
results.
\citet{Senarathne+etal:2020:Laplace_adaptive_design} use PSIS to
stabilize a Laplace-based sequential Monte Carlo algorithm for adaptive
design of experiments.

It can also be expected that variance of particle filter estimates in
filtering and smoothing problems could be reduced with Pareto
smoothing of the weights and with improved regeneration control by
using Pareto $\hat{k}$ diagnostic. Analyzing the effect of these in
different population and sequential Monte Carlo and particle filter
algorithms and applications would be worthy of further study.

\section*{Acknowledgments}

We thank Juho Piironen for help with R implementation, Tuomas Sivula for help
with Python implementation, Viljami Aittom{\"a}ki for help with protein
expression data, Michael Riis Andersen, Seth Axen, Ozan Ad{\i}g\"{u}zel, Srikanth Cadicherla, Noa Kallioinen, Finn Lindgren, Shira Mitchell, and
anonymous reviewers for helpful comments,
and the Research Council of Finland (grant 313122), Research Council of Finland Flagship Programme Finnish Center for Artificial Intelligence (FCAI), Alfred P. Sloan Foundation, U.S. National Science
Foundation, Institute for Education Sciences, Office of Naval Research, Natural Sciences and Engineering Research Council of Canada, and Canadian Research Chair programme for
partial support of this research.

\appendix

\section{Scaling of Distribution of Mean of Truncated Means}
\label{sec:truncated_mean}

PSIS replaces the $M$ largest weights with the expected order statistics of the Pareto distribution. The
generalized central limit theorem states that the largest expected
order statistic scales as $S^k$ \citep[e.g.,][p.\ 138]{Bouchaud+Georges:1990}, and we can model the distribution of the mean of the expected order statistics as mean of truncated Pareto variables. \citet{Zaliapin+etal:2005} provide equations for mean and variance of truncated Pareto distribution:
\begin{align*}
&\mu_{y}=\E\left(X_{1} \mid X_{1} \leq y\right)= \begin{cases}\frac{1}{k-1}\left(y^{1-{1/k}}-1\right) /\left(1-y^{-{1/k}}\right), & \qquad k \neq 1, \\
\log (y) /\left(1-y^{-1}\right), & \qquad k=1,\end{cases} \\
&\sigma_{y}^{2}=\Var\left(X_{1} \mid X_{1} \leq y\right)= \begin{cases}\frac{1}{2k-1}\left(y^{2-{1/k}}-1\right) /\left(1-y^{-{1/k}}\right)-\mu_{y}^{2}, & k \neq 0.5, \\
2 \log (y) /\left(1-y^{-2}\right)-\mu_{y}^{2}, & k=0.5.\end{cases}
\end{align*}

We set the truncation point to the largest expected order statistic,
$y=S^k$, and get
\begin{align*}
&\mu_{y}=\E\left(X_{1} \mid X_{1} \leq S^k\right)= \begin{cases}\frac{1}{k-1}\left(S^{k-1}-1\right) /\left(1-S^{-1}\right), & \qquad k \neq 1, \\
\frac{1}{2}\log(S) /\left(1-S^{-1}\right), & \qquad k=1,\end{cases} \\
&\sigma_{y}^{2}=\Var\left(X_{1} \mid X_{1} \leq S^k\right)= \begin{cases}\frac{1}{2k-1}\left(S^{2k-1}-1\right) /\left(1-S^{-1}\right)-\mu_{y}^{2}, & k \neq 0.5, \\
\log (S) /\left(1-S^{-1}\right)-\mu_{y}^{2}, & k=0.5.\end{cases}
\end{align*}

When $S$ is big and $k\in (0, 0.5-0.5/\log_{10}(S))$, the RMSE is dominated by the variance, $\sigma_y^2$ is dominated by $\E(X_1^2)$, and
\begin{align*}
  \left(S^{2k-1} -1 \right) /\left(S^{-1} - 1 \right) \in (0.9, 1).
\end{align*}
Thus the effect of $k$ is small for $\sigma_y^2$, and the standard
deviation of mean of Pareto truncated variables then scales
approximately as $S^{-1/2}$ matching CLT scaling (the standard
deviation can still depend on $k$).  Interestingly, the above
threshold ($0.5-0.5/\log_{10}(S)$) is equal to half of the max $\hat{k}$ threshold given $S$ (see
Section \ref{sec:S-k-threshold}). The scale rate change around this
threshold is smooth and threshold is not describing a sharp phase
transition.

When $S$ is big and $k = 0.5$, the variance dominates bias, the RMSE
is dominated by the variance, $\sigma_y^2$ is dominated by $\E(X_1^2)$, and
\begin{align*}
\log (S) /\left(1-S^{-1}\right) \approx \log(S).
\end{align*}
Thus the standard deviation of mean of Pareto truncated variables
scales approximately as $(S/\log(S))^{-1/2}$, matching the GCLT
scaling (at $k=0.5$).

When $S$ is big and $k\in (0.5 + 0.5/\log_{10}(S), 1)$, we need to
consider the effect of both bias and variance for RMSE. Bias is
\begin{align*}
  \mu(\left(1-S^{k-1}\right) /\left(1-S^{-1}\right) - 1) & \approx \mu S^{k-1}.
\end{align*}
$\sigma_y^2$ is dominated by $\E(X_1^2)$, and
\begin{align*}
  \left(S^{2k-1}-1\right) /\left(1-S^{-1}\right) & \approx S^{2k-1}.
\end{align*}
The standard deviation of mean of truncated Pareto variables scales
then approximately as $(S^{2k-1})^{1/2}S^{-1/2}=S^{k-1}$.  Both
bias and standard deviation of mean scale as $S^{k-1}$, so RMSE
scales also as $S^{k-1}$.  The additional terms, which depend on $k$,
make the bias to grow faster than the standard deviation, and thus
eventually when $k$ is big enough---approximately $0.7$---RMSE is
dominated by the bias; see also Section \ref{sec:PSIS_RMSE}.

The behavior of scaling of mean of truncated Pareto variables varies
smoothly with $k$, but approximately we have CLT scaling $S^{-1/2}$
when $k\in (0, 0.5 - 0.5/\log_{10}(S))$ and GCLT scaling $S^{k-1}$ when
$k\in (0.5 + 0.5/\log_{10}(S), 1)$, between these the scaling changes
smoothly and monotonically so that it is $(S/\log(S))^{-1/2}$ when
$k=0.5$.  When $S$ increases, the range
$k\in(0.5-0.5/\log_{10}(S), 0.5 +0.5/\log_{10}(S))$ gets shorter.

The above equations are approximate as for simplicity we have dropped minor terms,  and they do not take into account 1) the reduced variance of PSIS due to replacing $M$ largest weights with expected order statistics, and 2) the difference in bias between analytic integral and use of (approximated) expected order statistics. Based on the experiments these differences do not affect the order of the scaling.

For self-normalized truncated importance sampling (TIS), the truncation point is $y=\bar{r}S^{1/2}$, where $\bar{r}=\frac{1}{S}\sum_{s=1}^Sr_s$ \citep[][Appendix B]{Ionides:2008}. This truncation level tends to be lower than truncation in PSIS, which reduces the variance, but increases bias. Analyzing the effect of $\bar{r}$ is complicated as it has infinite variance. When $k < 1$ the truncated mean is
\begin{align*}
  \frac{1}{k-1}\left(cS^{1/2-1/(2k)}-1\right)/\left(1-cS^{-1/(2k)}\right).
\end{align*}
When $S$ is big and $0.5 \ll k < 1$,
\begin{align*}
\left(cS^{1/2-1/(2k)}-1\right)/\left(1-cS^{-1/(2k)}\right) \approx \left(cS^{1/2-1/(2k)}-1\right),
\end{align*} 
and TIS bias scales as $cS^{1/2-1/(2k)}$ and thus grows faster than
PSIS bias. Here $c$ includes the effect of $\bar{r}$ which depends on the observed ratios. Specifically, TIS truncation fails sometimes when there is one extremely large weight that causes the truncation level to rise so high that other large weights are not truncated (demonstrated in the examples).

\begin{figure}[t]
  \centering
  \includegraphics[height=2.8in]{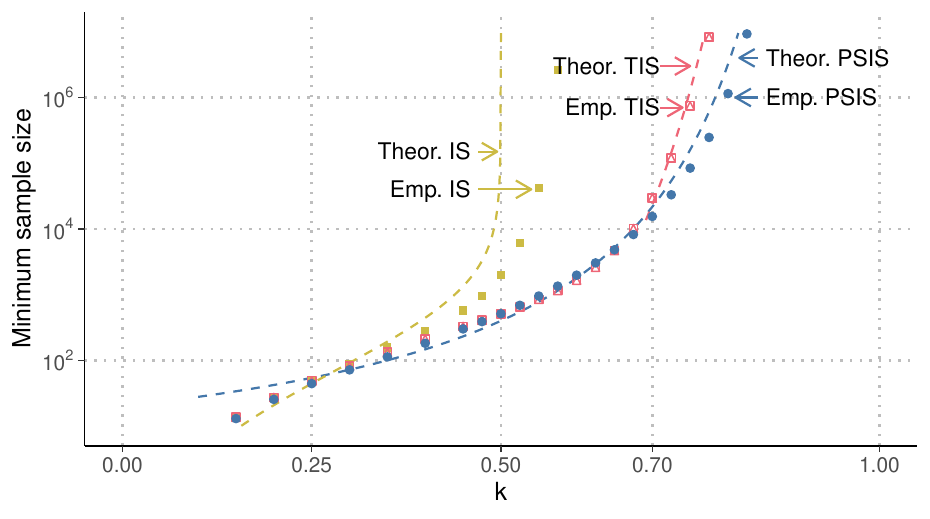}
   \vspace{-.25\baselineskip}
  \caption{Theoretical (dashed lines) and empirical (dots) minimum sample size to control RMSE for {\color{matblue}PSIS}, {\color{matred}TIS}, and {\color{matyellow}IS} as a function of $k$. Empirical results are obtained with 10\,000 repeated simulations, using grid of $k$ and $S$ values and interpolation, to find the sample sizes needed to obtain RMSE of $0.05$.}
  \label{fig:RMSE_minimum_sample_size_PSIS_TIS_IS}
\end{figure}
Figure \ref{fig:RMSE_minimum_sample_size_PSIS_TIS_IS} shows the theoretical and empirical minimum sample sizes for PSIS, TIS, and IS. Theoretically IS has infinite variance when $k>0.5$, but the empirical RMSE result for IS is optimistic as the error distribution has a thick tail, and with finite number of repeated simulations that tail is not well explored. This optimism means also that when $k$ is close to $0.5$ or larger, IS with finite number of draws is not unbiased. The empirical behavior of IS RMSE depends on the sample size $S$, but also on the number of repeated simulations. TIS has similar RMSE as PSIS until approximately $k>0.65$, but the RMSE of TIS is dominated by the bias much earlier and thus variance based MCSE estimates are less useful for TIS than for PSIS. See more discussion in Section \ref{sec:PSIS_RMSE}.

\section{Relative Convergence Rates}
\label{sec:rate_appendix}

When $k < 0.5-0.5/\log_{10}(S)$, the root mean squared error of PSIS scales approximately as $S^{-1/2}$ as under CLT. If we write this as $(S^{-1/2})^\alpha, \alpha=1$, we can then consider relative convergence rates $\alpha$ compared to the CLT rate. For $k\in (0.5+0.5/\log_{10}(S), 1)$, $(S^{-1/2})^\alpha=S^{(k-1)}$ is solved by $\alpha=2-2k$. When $k=0.5$,  
\begin{align*}
  (S^{-1/2})^\alpha & = (S/\log(S))^{-1/2}&  &  \\ 
  S^\alpha & = (S/\log(S)) & \vert\ & \log &\\ 
  \alpha\log(S)  & = \log(S/\log(S)) & \vert\ & \frac{d}{dS}& \\ 
   \alpha/S & = (1-1/\log(S))/S & & \\ 
   \alpha & = 1-1/\log(S). & & 
\end{align*}
The above simple equations help to provide insight to the behavior of relative
convergence rate for $k \ll 0.5$, $k=0.5$, and $k \gg 0.5$. We can also
derive more accurate smooth approximation from $\E(X_1^2)$ of
truncated Pareto mean (see Appendix
\ref{sec:truncated_mean}) and get relative convergence rate
\begin{align*}
  \begin{cases}
    \frac{2(k-1)S^{(2k+1)}+(1-2k)S^{(2k)}+S^2}{(S-1)(S-S^{(2k)})},  & k \neq 0.5, \\
    1-1/\log(S), & k=0.5,
  \end{cases}
\end{align*}
which can be used as part of computer assisted diagnostics.

\begin{figure}[t]
  \centering
  \includegraphics[width=\textwidth]{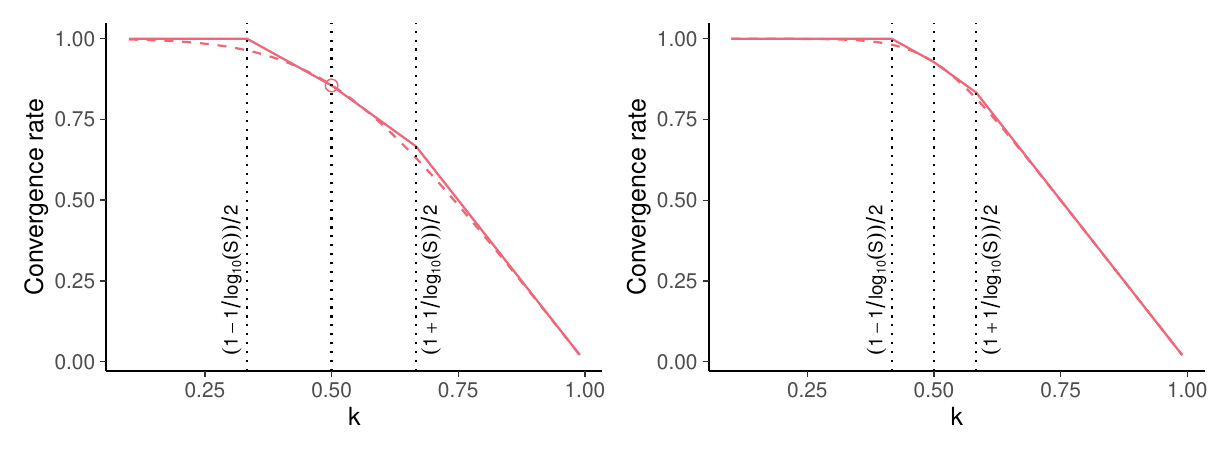}
   \vspace{-1.25\baselineskip}
   \caption{Convergence rate as a function of $k$ and $S$. {\color{matred} Red solid line} shows a piecewise linear approximation and {\color{matred} Red dashed line} shows a smooth approximation. Both approximations are based on properties of distribution of mean of truncated Pareto variables (see text). Left subplot shows the result with $S=10^3$  and the right plot shows the result with $S=10^6$. The dotted lines show the approximate thresholds used for the piecewise linear approximation. Empirical results are obtained from 10\,000 repeated simulations.}
  \label{fig:convergence_rate_piecewise_vs_smooth}
\end{figure}
Figure \ref{fig:convergence_rate_piecewise_vs_smooth} shows piecewise linear and smooth approximations of the
relative convergence rate when $S=10^3$ and $S=10^6$. Figure
\ref{fig:relative_efficiency} in Section \ref{sec:rate} shows the
relative convergence rate approximation (smooth) and empirical results when
$S=10^4$. These convergence rate approximations match well the empirical results in Figures \ref{fig:exptest_convrateqq_combined}, \ref{fig:xoff_convrateqq_combined}, and \ref{fig:dimxoff_convrateqq_combined}, except in the fourth subplot of Figure \ref{fig:xoff_convrateqq_combined}. In that case, the actual convergence rate for first and second moment is better than predicted by $h$-specific $\hat{k}_{h}$. This is explained by having a bounded ratio distribution, and initially not observing ratios near the bound, but when the sample size increases the bound starts to have an effect and convergence is faster than initially predicted (see also Section \ref{sec:bounded_ratios}).

When $k=0.5$ and $S \rightarrow \infty$, the convergence rate approaches $1$, but with any finite $S$ the convergence rate is less than $1$.

\section{Proof of Theorem \ref{thm:convergence}} \label{appendix:proof}

We can write the PSIS estimator as a Winsorized version of the truncated
estimator of \citet{Ionides:2008} plus an extra bias correction term:
$$
I^S_h =
  \frac{1}{S} \sum_{s=1}^{S}\left(r(\theta_s) \wedge r_{(S-M+1):S}\right)h(\theta_s)
  \,+\, \frac{1}{S}\sum_{j=1}^M w^{+}_j h(\theta_{S-M+j }).
$$
For truncated importance sampling, the number of draws, $M$, that exceed the threshold is random while the threshold is fixed. For the Winsorized version the number of exceedences is fixed but the threshold is random.
The threshold is the $M$th largest value of $r_s$. We denote this using order
statistics notation: we reorder the sample so that 
$$
r_{1:S} \leq r_{2:S}\leq \dots r_{S:S}.
$$ 
With this notation, the threshold is $T = r_{S-M+1:S}$ and the Winsorized importance sampler (WIS)
is 
$$
I^S_\text{WIS} = \frac{1}{S}\sum_{s = 1}^{S-M} h_{s:S}r_{s:S} + \frac{r_{S-M+1:S}}{S}\sum_{s=S-M+1}^S h_{s:S},
$$
where $(r_{s:S}, h_{s:S})$ are the $(r_s, h_s)$ pairs ordered so that $r_{1:S} \leq r_{2:S}\leq \dots \leq r_{S:S}$.

Conditioning on the event $r_{S-M+1:S} = T$ yields,
$$
\mathbb{E}\left(I_\text{WIS}^S \mid r_{S-M+1:S} = T\right) = \left(1 - \frac{M}{S}\right)\mathbb{E}(RH \mid R < T) + \frac{MT}{S} \mathbb{E}(H \mid R \geq T).
$$
From this, the bias, conditional on $r_{S-M+1:S} = T$, is 
\begin{multline*}
\left|I - \mathbb{E}\left(I_\text{WIS}^S \mid r_{S-M+1:S} = T\right)\right| =\left|\left(\Pr(R < T) - \left(1 - \frac{M}{S}\right)\right)\mathbb{E}(RH \mid R < T) \right.\\ 
\left.+ \left[\Pr(R \geq T) - \frac{M}{S}\right] \mathbb{E}(H(R - T) \mid R \geq T)\right|.
\end{multline*}
Using the triangle inequality
$$
\left|\mathbb{E}(RH \mid R > T)\right| \leq \frac{\mathbb{E}(R |H| 1(R<T))}{\Pr(R <T)} \leq \frac{\|h\|_{L^1(p)}}{\Pr(R  <T)},
$$
we see that the first term in the bias can be bounded if we can bound the relative error,
$$
\mathbb{E}\left|1 - \frac{1- M/S}{\Pr(R < r_{S-M+1:S})}\right|.
$$
To attack this expectation, we use the following lemma that derives the distribution of the relative error we incur when approximating $\Pr(R > r_{S-M+1:S})$ with $1 - M/S$.
\begin{lemma}
Let  $x_s$, $s= 1, \ldots S$ be an iid sample from $F_X$, and let $r\in [0, S]$
be an integer. Then, for any $p\in (0,1)$,
$$
\frac{p}{F_X(x_{r:S})} -p \stackrel{d}{=} \frac{p(S-r+1)}{r} \mathcal{F},
$$
and 
$$
\frac{1-p}{1- F_x/(x_{r:S})} - (1-p) \stackrel{d}{=} \frac{k(1-p)}{S-r+1}\mathcal{F}^{-1},
$$
where $\mathcal{F}$ is an F-distributed random variable with parameters $(2(S-k+1), 2k)$.
\end{lemma}
\begin{proof}
For any $t\geq 0$, 
\begin{align*}
\Pr\left(\frac{p}{F_X(x_{k:S})} - p \leq t\right) &=\Pr\left(p - pF_X(x_{k:S}) \leq tF_X(x_{k:S})\right) \\
&= \Pr\left(p  \leq (t+p)F_X(x_{k:S})\right) \\
&=\Pr\left(F_X(x_{k:S}) \geq \frac{p}{p+t}\right)\\
&= \Pr\left(x_{k:S} \geq F_X^{-1}\left(\frac{p}{p+t}\right)\right)\\
&= 1- I_{\frac{p}{p+t}}(k, S-k+1) \\
&= I_{\frac{t}{p+t}}(S-k+1, k),
\end{align*}
where $I_p(a,b)$ is the incomplete beta function; see equation 2.1.5
in \citet{David+Nagaraja:2004}.
\end{proof}

It follows then that 
$$
\left|1 - \frac{1-M/S}{R(r_{S-M+1})} \right| \stackrel{d}= \left|\frac{M}{S} -  \frac{M(S-M)}{S(S-M-1)}\mathcal{F}\right| \leq \frac{M}{S} +  \frac{M(S-M)}{S(S-M-1)} \mathcal{F},
$$
where $\mathcal{F}$ has an F-distribution with $(M, S-M+1)$ degrees of freedom.
As $\mathbb{E}(\mathcal{F}) = 1 + 1/(S-M-1)$, it follows that this term goes to zero
as long as $M = o(S)$. This shows that the first term in the bias goes to zero.

The second term in the bias is 
$$
\left(\Pr(R \geq T) - \frac{M}{S}\right) \mathbb{E}(H(R - T) \mid R \geq T).
$$
As before, we can write this as 
$$
\left(1 - \frac{M/S}{1-R(T)}\right)|\mathbb{E}(H(R - T) 1_{R \geq T})| \leq \left|1 - \frac{M/S}{1-R(T)}\right|\|h\|_{L^1(p)}.
$$
By our lemma, we know that the distribution of the term in the absolute value 
when $T = r_{S-M+1}$ is the same as 
$$
1-\frac{M}{S} -\left(1 - \frac{M}{S} + \frac{1}{S}\right)\mathcal{F} = (\mu_F-\mathcal{F})  +\frac{M}{S}(\mathcal{F}-\mu_F) - \frac{1}{S}\mathcal{F} +  \frac{1}{M-1}\left(\frac{M}{S} - 1\right),
$$
where $\mathcal{F} \sim \text{F}_{2(S-M+1), 2M}$, which has mean $\mu_F = 1+(M-1)^{-1}$ and variance 
$$
\sigma^2_F = \frac{M^2S}{(S-M+1)(M-1)^2(M-2)} = \frac{1}{M}\left(1 + \mathcal{O}(M^{-1} + MS^{-1})\right).
$$
From Jensen's inequality, we get $$
\mathbb{E}(|\mathcal{F} - \mu_F|) \leq \sigma_F = M^{-1/2}(1 + o(1)).
$$
It follows that 
\begin{multline*}
\mathbb{E}\left|1 - \frac{M/S}{1-R(r_{S-M+1:S})}\right| \leq\\ M^{-1/2}(1+o(1))M^{1/2}S^{-1}(1 + o(1)) + S^{-1}(1+ o(1)) + (M-1)^{-1}(1+o(1)),
\end{multline*}
and so we get vanishing bias as long as $M\rightarrow \infty$ and $M/S \rightarrow 0$.

\begin{theorem}
Let $\theta_s$, $s = 1,\ldots, S$ be an iid sample from $G$ and let $r_s = r(\theta_s) \sim R$. Assume that
\begin{enumerate}
\item $R$ is absolutely continuous,
\item $M  \rightarrow \infty$ and $S^{-1}M \rightarrow 0$, and
\item $h \in L^1(p)$.
\end{enumerate}
Then the Winsorized importance sampling converges in $L^1$ and is
asymptotically unbiased.
\end{theorem}

Assuming that $R - T \mid R\geq T$ is in 
the domain of attraction of a generalized Pareto distribution with shape parameter $k$.
A sufficient condition, due to von Mises, is that 
$$
\lim_{r\rightarrow \infty} \frac{r R'(r)}{1-R(r)} = \frac{1}{k}.
$$
This is basically a regularity condition at infinity.
For example if $1-R(r)$ is regularly varying at infinity and $R'(r)$ is, eventually, monotone decreasing, then this condition holds.

The von Mises condition is natural here, as \citet{Falk+Marohn:1993} show that the relative error from approximating the tail of $R$ by a generalized Pareto density is the 
same as the relative error in the von Mises condition.  That is, if 
$$
\frac{rR'(r)}{1-R(r)} = \frac{1}{k}(1 + \mathcal{O}(r^{-\alpha}))
$$
then 
$$
R'(r) = c w(cr - d)(1 + \mathcal{O}(r^{-\alpha})),
$$
where $c$ and $d$ are constants and $w$ is the density of a generalized Pareto distribution.

Under those two assumptions, we can swap out the density of $(R-T)\mid R>T$ with
its asymptotic approximation and get that, conditional on $T=  r_{S-M+1:S}$, 
$$
\mathbb{E}(H(R-T) \mid R>T) = (k-1)^{-1}T.
$$
Hence, the second term in the bias goes to zero if 
$$
\mathbb{E}\left(r_{S-M+1:S}\left(1 - R(r_{s-M+1:S}) - \frac{M}{S}\right)\right)
$$
goes to zero.

This helps us understand that even if 
a distribution doesn't have finite moments, away from the extremes its order 
statistics always do. This means that we can use Cauchy-Schwartz to get 
\begin{multline*}
  \left|\mathbb{E}\left(r_{S-M+1:S}\left(1 - R(r_{s-M+1:S}) - \frac{M}{S}\right)\right)\right| \leq\\
  \mathbb{E}\left(r_{S-M+1:S}^2\right)^{1/2}\mathbb{E}\left(\left(1 - R(r_{s-M+1:S}) - \frac{M}{S}\right)^2\right)^{1/2}.
\end{multline*}
Leaning into a result by \citet{Bickel:1967} and fixing a typo there, we get that 
$$
\mathbb{E}(r_{k:M}^2) \leq C k\begin{pmatrix} S \\ k\end{pmatrix} \int_0^1 t^{k-2-1}(1-t)^{S-k-2}\,dt.
$$
Noting 
that $B(n,m) = \Gamma(n)\Gamma(m)/\Gamma(n+m)$, we can rewrite the upper bound in
terms of the beta function to  get 
$$
\mathbb{E}(r_{k:M}^2) \leq C \frac{\Gamma(S+1)}{\Gamma(S-3)} \frac{\Gamma(k-2)}{\Gamma(k+1)}\frac{\Gamma(S-k-1)}{\Gamma(S-k+1)}.
$$
To show that this doesn't grow too quickly, we use the identity 
$$
\frac{\Gamma(x + a)}{\Gamma(x + b)} \propto x^{a-b}(1 + \mathcal{O}(x^{-1})),
$$
from which it follows that 
$$
\mathbb{E}(r_{k:M}^2) \leq C S^4k^{-3}(S-k)^{-2}(1+ \mathcal{O}(S^{-1}))(1+ \mathcal{O}(k^{-1}))(1+ \mathcal{O}((S+k)^{-1})).
$$
In this case, we are interested in $k = S-M+1$, so 
\begin{multline*}
  \mathbb{E}(r_{k:M}^2) \leq \\
  C S^4S^{-3}M^{-2}(1 - M/S + 1/S)^{-3}(1 - 1/M)^{-2}(1+ \mathcal{O}(S^{-1}))(1+ \mathcal{O}(S^{-1}))(1+ \mathcal{O}(M^{-1})).
\end{multline*}

Hence $\mathbb{E}(r_{k:M}^2) = \mathcal{O}(SM^{-2})$. This is increasing in 
$S$, but we will see that it is not going up too fast.

For the second half, we attack 
\begin{multline*}
  \mathbb{E}\left(\left(1 - R(r_{s-M+1:S}) - \frac{M}{S}\right)^2\right) = \\
  \mathbb{E}\left(\left(1 - R(r_{s-M+1:S})\right)^2 - 2\left(1 - R(r_{s-M+1:S})\right)\frac{M}{S} +\left(\frac{M}{S}\right)^2\right).
\end{multline*}
A standard result from extreme value theory is that $R(r_{k:S})$ has the same 
distribution as the $k$th order statistics from a sample of $S$ independent $\text{uniform}(0,1)$ 
random variables. Hence,
$$
R(r_{S-M+1:S}) \sim \text{beta}(S-M+1, M).
$$
If follows that 
$$
\mathbb{E}(1- R(r_{S-M+1:S})) = \frac{M}{S+1} = \frac{M}{S}\frac{1}{1+S^{-1}}
$$
and 
$$
\mathbb{E}((1- R(r_{S-M+1:S}))^2) = \frac{M(M+1)}{(S+1)(S+2)} = \frac{M^2}{S^2}\left(\frac{1 + M^{-1}}{1 + 3S^{-1} + 2S^{-2}}\right).
$$
Adding these and doing asymptotic expansions yields,
$$
\mathbb{E}\left(\left(1 - R(r_{s-M+1:S}) - \frac{M}{S}\right)^2\right) = \frac{M^2}{S^2} + \mathcal{O}\left(\frac{M}{S^2}\right),
$$
which goes to  zero like $\mathcal{O}(S^{-1})$ if $M = \mathcal{O}(S^{1/2})$.

We can multiply this rate together and get that the second term in the bias 
is bounded above by 
$$
\left[\left(\frac{S}{M^2} (1 + \mathcal{O}(M^{-1} + MS^{-1}))\right)\left(\frac{M^2}{S^2} (1 + \mathcal{O}(M^{-1} + MS^{-1})\right)\right]^{1/2} = S^{-1/2}(1 + o(1)).
$$
Putting all of this together we have proved the following corollary.

\begin{corollary}
Let $\theta_s$, $s = 1,\ldots, S$ be an iid sample from $G$ and let $r_s = r(\theta_s) \sim R$. Assume that
\begin{enumerate}
\item $R$ is absolutely continuous and satisfies the von Mises condition 
$$
\frac{rR'(r)}{1-R(r)} = \frac{1}{k}(1 +\mathcal{O}(r^{-1})),
$$
\item $M  = o(S)$, and
  
\item $h$ is bounded.
\end{enumerate}
Then Winsorized importance sampling converges in $L^1$ with rate of, at most, $\mathcal{O}(MS^{-1} + S^{-1/2})$,
which is balanced when $M = \mathcal{O}(S^{1/2})$. Hence, WIS is $\sqrt{n}$-consistent.
\end{corollary}

To tackle the variance part we use
$$
\mathbb{V}\left(I_\text{WIS}^S\right) \leq \mathbb{E}\left([I_\text{WIS}^S]^2\right),
$$
noting that we can write $I_\text{WIS}^S$ compactly as 
$$
I_\text{WIS}^S = \frac{1}{S}\sum_{s=1}^S h(\theta_s)\min\{r(\theta_s), r_{S-M+1:S}\}.
$$
Hence,
\begin{align*}
\mathbb{E}\left([I_\text{WIS}^S]^2\right) &= \mathbb{E}_{T\sim r_{S-M+1:S}}\left[\mathbb{E}\left([I_\text{WIS}^S]^2 \mid r_{S-M+1:S} = T\right)\right]\\
&=\frac{1}{S^2}\mathbb{E}_{T\sim r_{S-M+1:S}}\left[\mathbb{E}\left(H^2 \min\{R^2,T^2\} \mid r_{S-M+1:S} = T\right)\right]\\
&\leq\frac{1}{S^2}\mathbb{E}_{T\sim r_{S-M+1:S}}\left[\mathbb{E}\left(RTH^2 \mid r_{S-M+1:S} = T\right)\right] \\ 
&\leq\frac{1}{S^2}\mathbb{E}_{T\sim r_{S-M+1:S}}\left[T\|h\|_{L^2(p)}^2\right] ,
\end{align*}
which goes to zero as long as $\mathbb{E}(r_{S-M+1:S}) = o(S^2)$.

\citet{Bickel:1967} shows that,
noting that $\mathbb{E}(R) < \infty$, 
$$
\mathbb{E}(r_{S-M+1:S}) \leq C (S-M+1)\frac{\Gamma(S+1)\Gamma(S-M+1-1)\Gamma(M)}{\Gamma(S-M+1+1)\Gamma(M+1)\Gamma(S-1)} = \frac{S}{M}(1 + o(1)),
$$
and so the variance is bounded.

The previous argument shows that the variance is $\mathcal{O}(M^{-1}S^{-1})$. We
can refine that if we assume the von Mises 
condition hold.  In that case we know that $R(r) = 1- cr^{-1/k} + o(1)$ as $r\rightarrow \infty$  and therefore 
\begin{align*}
R\left(R^{-1}\left(1-\frac{M}{S}\right)\right) &= 1-\frac{M}{S+1}\\
1 - cR^{-1}\left(1-\frac{M}{S+1}\right)^{-1/k}(1+o(1)) &= 1- \frac{M}{S+1} \\
R^{-1}\left(1-\frac{M}{S+1}\right) &= c^{-k}\left(\frac{M}{S+1}\right)^{-k}(1 + o(1)).
\end{align*}
\citet{Bickel:1967} shows that $\mathbb{E}(r_{k:S}) = R^{-1}(1-M/(S+1)) + o(1)$, so combining
this with the previous result gives a variance of $\mathcal{O}((M/S)^{k-2})$. If 
we take $M =\mathcal{O}(S^{1/2})$, this gives $\mathcal{S}^{k/2-1}$, which is 
smaller than the previous bound for $k<1$. 
Hence the variance goes to zero and we have proved the following theorem.

\begin{theorem}
Let $\theta_s$, $s = 1,\ldots, S$ be an iid sample from $G$ and let $r_s = r(\theta_s) \sim R$. Assume that
\begin{enumerate}
\item $R$ is absolutely continuous,
\item $M \rightarrow \infty$ and $M^{-1}S \rightarrow 0$, and
\item $h \in L^2(p)$.
\end{enumerate}

The variance in Winsorized importance sampling is at most $\mathcal{O}(M^{-1}S)$.
\end{theorem}

Next we look at the PSIS bias correction.
Essentially, it works by noting that approximating 
$$
(1-R(r_{S-M+1:S}))\mathbb{E}(HR \mid R>r_{S-M+1:S}) \approx \frac{1}{S}\sum_{m=1}^M w_m h_{S-M+m:S},
$$
where $w_m$ is the median $m$th order statistic in an iid sample of $M$ generalized Pareto random variables with tail parameters fitted to the distribution.
This is equivalent to a quadrature rule. To see that, we can write 
$$
\mathbb{E}(HR \mid R>T) = \mathbb{E}(R\, \mathbb{E}(H \mid R)).
$$
If we approximate the distribution of $R > T$ by 
$$
\tilde{R}_\text{PSIS}(r) = \frac{1}{M}\sum_{m=1}^M 1( w_m<r)
$$
and approximate the conditional probability by 
$$
\Pr(H < h\mid R = w_m) \approx 1(h_{S-M+m:S}< h),
$$
then the convergence and vanishing variance
still holds. To see this, we note that 
$$
w_m = r_{S-M+1}  + k^{-1}\sigma\left(\left(1-\frac{j-1/2}{M}\right)^{-k} -1\right).
$$
So we are just reweighting our tail $H$ samples by 
$$
1 + \frac{\sigma}{kr_{S-M+1:S}}\left(\left(1-\frac{j-1/2}{M}\right)^{-k} -1\right).
$$

Recalling that when $R(r) = 1- cr^{-1/k}(1+ o(1))$, we had $\sigma = \mathcal{O}(r_{S-M+1:S})$,
this term is at most $\mathcal{O}(1 + M^{-k})$. This will not trouble either of our convergence proofs.

This leads to the following modification of our previous result, which was stated in Section \ref{sec:asymptotic}, but repeated here for convenience.
\setcounter{theorem}{0}
\begin{theorem}
  Let $\theta_s$, $s = 1,\ldots, S$ be an iid sample from $G$ and let $r_s = r(\theta_s) \sim R$. Assume that
\begin{enumerate}
\item $R$ is absolutely continuous,
\item $M  = \mathcal{O}(S^{1/2})$,
\item $h \in L^2(p)$, and
\item $k$ and $\sigma$ are known with $\sigma = \mathcal{O}(r_{S-M+1:S})$.
\end{enumerate}
Then Pareto smoothed importance sampling converges in $L^1$ and its variance goes to zero and 
it is consistent and asymptotically unbiased.
\end{theorem}

In addition, we have the following corollary.
\begin{corollary}
  Assume further that
  \begin{enumerate}
  \item $R$ satisfies the von Mises condition
$$
\frac{rR'(r)}{1-R(r)} = \frac{1}{k}(1 +\mathcal{O}(r^{-1})), \text{and}
$$
\item $h$ is bounded.
  \end{enumerate}
Then the $L^1$ convergence occurs at a rate of, at most, $\mathcal{O}(S^{-1/2})$. 
Furthermore, the variance of the PSIS estimator goes to zero at least as fast as  $\mathcal{O}(S^{k/2-1})$. 

Hence, under these additional conditions PSIS is $\sqrt{n}$-consistent.
\end{corollary}

\section{Simulated Examples}
\label{sec:examples}

In the following experiments, we know the true target value and
vary the proposal distribution. This allows us to study how $\hat{k}$
functions as a diagnostic and how bad the approximating distribution has to be
before the importance sampling estimates and corresponding variance based Monte Carlo standard error (MCSE) estimates defined in Section \ref{s:pareto}
break down. To diagnose the
performance with respect to the number of draws $S$, in each of the
examples we vary the number of draws from $S=10^2$ to $S=10^5$. We
examine the estimates for the normalization term (zeroth moment), which
also has a special role in the importance sampling examples, and
estimates for $\E(h(\theta))$, where $h(\theta)=\theta$ (first moment) or
$h(\theta)=\theta^2$ (second moment).
Although in these examples the normalization terms of both $p$ and $g$
are available, all experiments have been made assuming that
normalization terms are unknown and self-normalized importance
sampling is used.

\subsection{Exponential Target and Proposal} \label{sec:exponential_test}

In the first simulated example, the target and proposal distributions
are exponential distributions with mean parameter 1 and $1/\lambda$,
respectively. In this case, it is possible to compute the distribution
of the importance ratios in closed form.
When $\lambda>1$, the distribution of the importance ratios is a Pareto type I
distribution with shape parameter $k=1-1/\lambda$, and a generalized
Pareto distribution \eqref{pareto} with the same shape parameter $k=1-1/\lambda$,
 scale parameter $\sigma=k(1-k)$, and location parameter $u=\sigma/k$.

When estimating first and second moments, the distributions of
$\theta r(\theta)$ and $\theta^2 r(\theta)$ are not Pareto
distributed, but in the simulations the order of magnitude for the
sample size to obtain small root mean square errors is similar.

\begin{figure}[tp]
    \hspace{17pt}\includegraphics[width=0.96\textwidth]{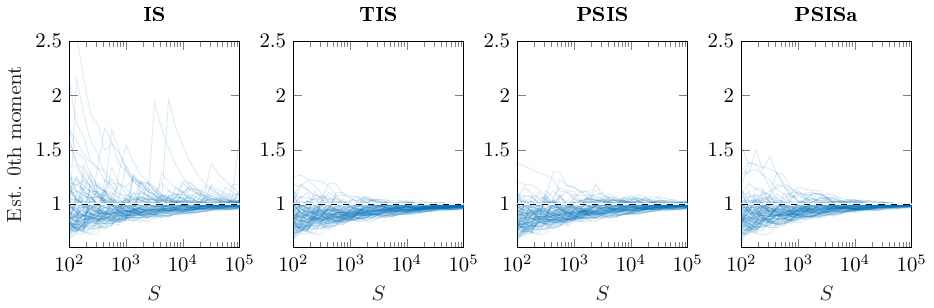}\\
    \includegraphics[width=0.98\textwidth]{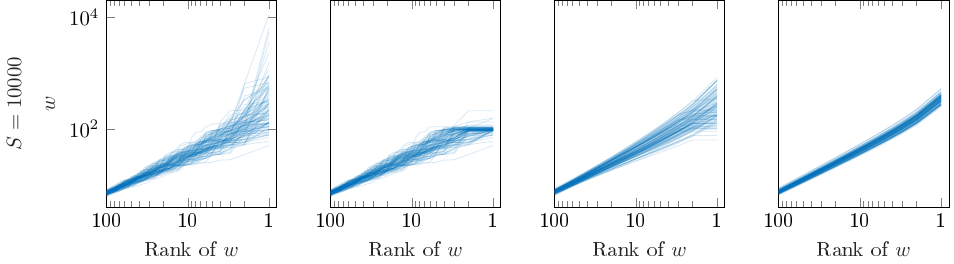}
\caption{Each graph has {\color{matblue} 100 blue lines}, and each line is the trajectory of
the estimated normalization term using different initial random seeds. The first
line shows the bias and variance of the different estimators when estimating the zeroth moment (normalization term)
as the MCMC sample size increases.  The second line shows the distribution of
the importance sampling weights when $S=10\,000$. The target and proposal
distributions are exponential distributions with mean parameters 1 and 1/3
respectively, and the distribution of ratios has infinite variance.
  }
  \label{fig:exptest}
\end{figure}
\begin{figure}[tp]
  \begin{center}
    \includegraphics[width=\textwidth]{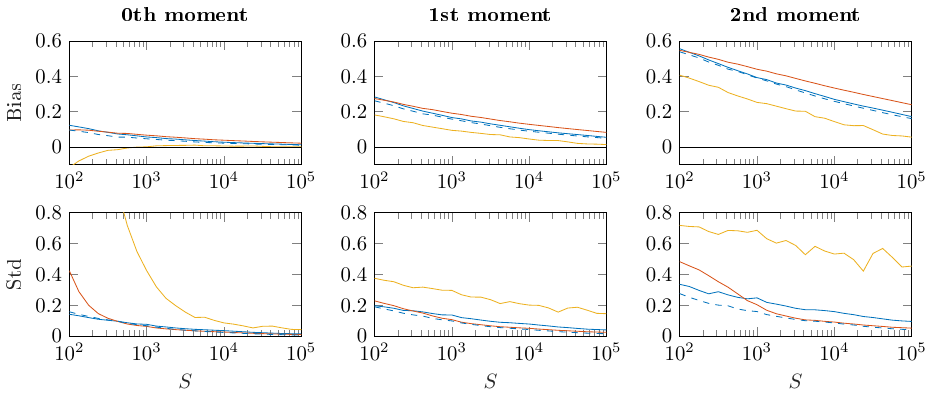}
  \end{center}
   \vspace{-\baselineskip}
  \caption{{\color{matyellow}IS
  is yellow \textbf{---}}, {\color{matred}TIS is red \textbf{---}}, {\color{matblue}PSIS is blue \textbf{---}} and {\color{matblue}PSISa is blue dashed - -}. Bias and standard deviation of the zeroth, first, and second moment
  estimates with respect to the number of draws $S$ computed from 1000
  simulations. Target and proposal distributions are exponential distributions
  with mean parameters 1 and 1/3 respectively, leading to $k\approx 0.66$.}
  \label{fig:exptest_biasvar}
\end{figure}

Figure~\ref{fig:exptest} shows 100 simulations of estimating
the normalization term (zeroth moment) using a proposal distribution with mean
parameter $1/3$, leading to $k\approx 0.66$ which illustrates the
typical behavior of IS, TIS and PSIS when $k\in(0.5, 0.7)$.
IS has high variability (empirical variance is finite due to the finite number of repeated simulations), while TIS and PSIS produce more stable estimates.
For the same 100 simulations, Figure~\ref{fig:exptest} shows
the 100 largest IS raw ratios and modified weights from TIS, PSIS, and
PSISa. PSISa uses all the raw ratios, that is $M=S$, to fit the
generalized Pareto distribution and estimate $\hat{k}$. IS has high
variability, TIS truncates the largest weights downwards, while PSIS
reduces the variability without biasing the largest weights. PSISa
uses more draws to estimate $\hat{k}$ and has smaller variability than
PSIS.

Figure~\ref{fig:exptest_biasvar} shows the bias and standard deviation of the
zeroth, first, and second moment estimates with respect to the number of draws
$S$ computed from 1000 simulations. This illustrates the typical
differences between the methods. IS has the smallest bias but
the largest deviation. PSISa has the smallest deviation,
and similar bias as PSIS. TIS has the largest bias and with
large $S$ similar deviation as PSISa.  PSIS has
similar bias as PSISa, but slightly larger deviation than PSISa.

Figure \ref{fig:minimum_sample_size} (in Section \ref{sec:PSIS_RMSE}) shows the average (over $10\,000$ simulations runs) sample size required for RMSE to be equal to $0.05$, when $\lambda$ is varied from $1.111$ to $20$, which correspond a range of $k$ from $0.1$ to $0.95$. The empirical result fits the theory in Section \ref{sec:reliable_psis} and Appendix \ref{sec:truncated_mean}.

\begin{figure}[tp]
  \begin{center}
    \includegraphics[width=\textwidth]{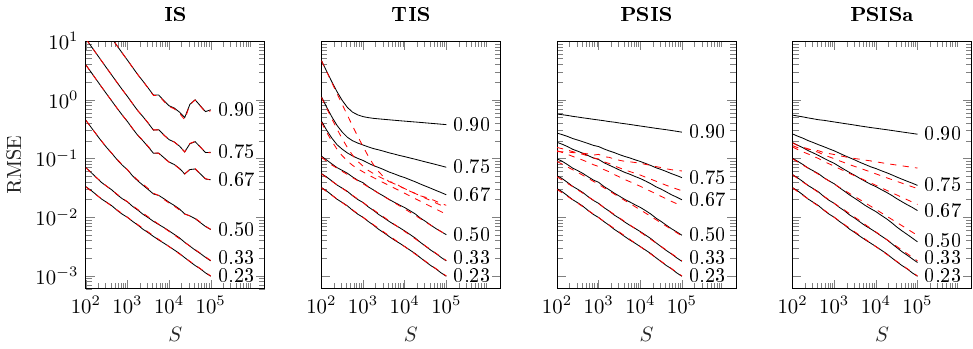}
  \end{center}
   \vspace{-\baselineskip}
  \caption{RMSE (black) and {\color{red}the mean of the MCSE estimates (red)} for the zeroth moment estimate. Target and proposal distributions are
  exponential with mean parameters 1 and $1/\lambda$ respectively,
  with $\lambda \in (1.3,1.5,2,3,4,10)$. The numbers at the end of black lines
  are average $\hat{k}$ values estimated when $S=10^5$.}
  \label{fig:exptest_error0}
\end{figure}

\begin{figure}[tp]
  \begin{center}
    \includegraphics[width=\textwidth]{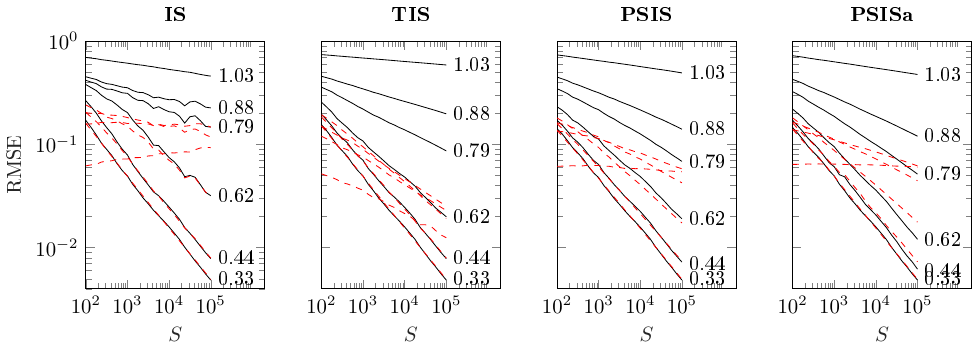}
  \end{center}
   \vspace{-\baselineskip}
  \caption{RMSE (black) and {\color{red}the mean of the MCSE estimates (red)} for the first moment estimate. Target and proposal distributions are
  exponential with mean parameters 1 and $1/\lambda$ respectively,
  with $\lambda \in (1.3,1.5,2,3,4,10)$. For each graph, the lines are ordered
  with low values of theta at bottom and high values at top, with high $\lambda$
  values leading to high RMSE and high $\hat{k}_h$. The numbers at the end of
  black lines are average $h$ specific $\hat{k}_h$ values estimated when
  $S=10^5$.}
  \label{fig:exptest_error1}
\end{figure}

\begin{figure}[tp]
  \begin{center}
    \includegraphics[width=\textwidth]{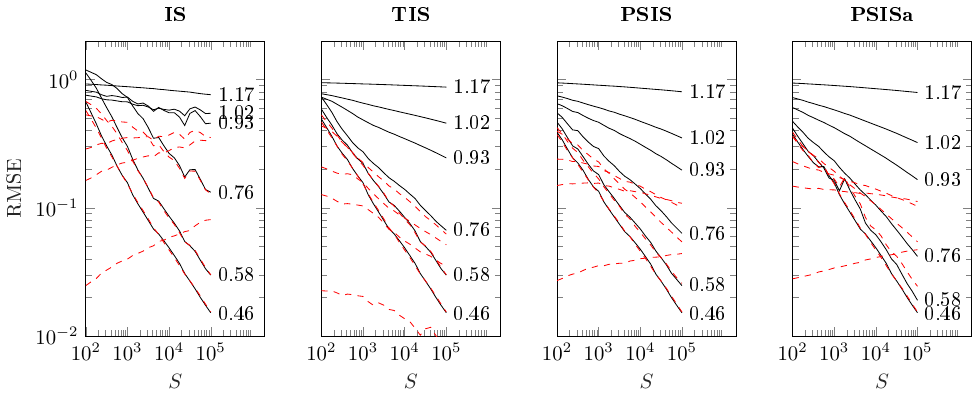}
  \end{center}
   \vspace{-\baselineskip}
  \caption{RMSE (black) and {\color{red}the mean of the MCSE estimates (red)} for the second moment estimate. Target and proposal distributions are
  exponential with mean parameters 1 and $1/\lambda$ respectively,
  with $\lambda \in (1.3,1.5,2,3,4,10)$. For each graph, the lines are ordered
  with low values of theta at bottom and high values at top, with high $\lambda$
  values leading to high RMSE and high $\hat{k}_h$. The numbers at the end of
  black lines are average $h$ specific $\hat{k}_h$ values estimated when
  $S=10^5$.}
  \label{fig:exptest_error2}
\end{figure}

Figures~\ref{fig:exptest_error0}, \ref{fig:exptest_error1}, and
\ref{fig:exptest_error2} show the RMSE and mean of the MCSE
estimates for zeroth, first and second moment estimates with varying
$\lambda \in (1.3,1.5,2,3,4,10)$.
The $h$-specific $\hat{k}_h$ and MCSE work well for first and second moment estimates. Simulations suggest that good results would be obtained
with $h(\theta)$ corresponding to higher moments, too.
All methods have impractical convergence rates in the case of larger $\hat{k}$ or $\hat{k}_h$. With
smaller $k$ or $\hat{k}_h$ values the errors are similar, but  already near
$k=0.5$, IS has larger RMSE. IS has finite empirical variance and RMSE with finite repeated simulations, but the tail shape of the error distribution matches the tail shape of the distributions of $r(\theta)$ and $r(\theta)h(\theta)$ revealing the infinite variance in the limit of increasing number of repeated simulations. TIS has larger RMSE than PSIS and PSISa for
large $\hat{k}$ values. There is not much difference in RMSE between PSIS and PSISa,
demonstrating the diminishing benefit of using a larger sample to estimate $\hat{k}$.
The empirical IS MCSE matches the empirical IS RMSE, but has high variance for
larger $\hat{k}$ (not shown here). The TIS MCSE estimates have negligible bias when $k<0.5$, but the error is underestimated for $k>0.5$ due to truncation bias dominating the error as discussed in Appendix \ref{sec:truncated_mean}.
The Monte Carlo estimates for PSIS and PSISa are useful for $k\lesssim 0.7$,
with PSISa yielding slightly more accurate error estimates.
The results from PSISa show that using larger $M$ (in PSISa $M=S$) would
give better estimates, but this works only if those $M$ largest weights are
well approximated by the generalized Pareto distribution.
\begin{figure}[tp]
  \begin{center}
    \includegraphics[width=\textwidth]{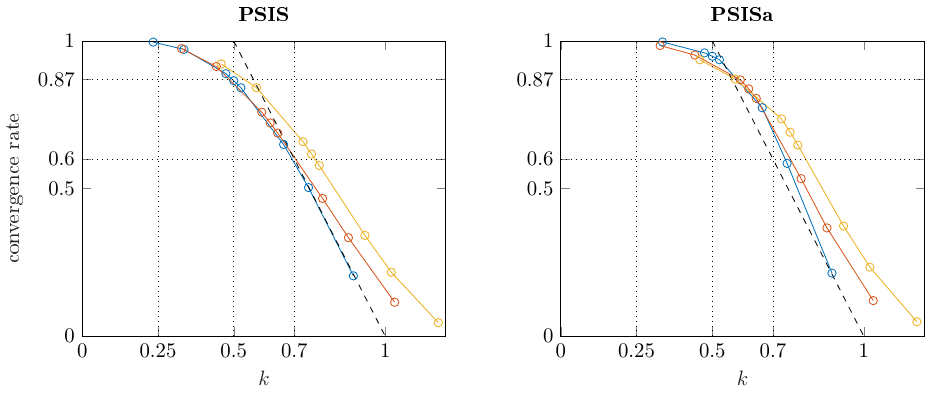}
  \end{center}
   \vspace{-1.25\baselineskip}
  \caption{Practical convergence rates for PSIS and PSISa estimating {\color{matblue}zeroth
  (blue)}, {\color{matred}first (red)}, and {\color{matyellow}second (yellow)} moments. Target and proposal distributions
  are exponential  with mean parameters 1 and $1/\lambda$
  respectively, with $\lambda \in (1.3,1.5,1.9,2,2.1,3,4,10)$. The circles in
  each line correspond to the results with different $\lambda$, with higher
  $\lambda$ values leading to lower convergence rates and higher $\hat{k}$
  values. The $\hat{k}$ estimates in case of first and second moments depend on $h$.
  Dotted horizontal lines at $0.87$ and $0.6$ show the typical convergence rates
  at $\hat{k}=0.5$ and $\hat{k}=0.7$ for several different experiments in this
  paper.}
  \label{fig:exptest_convrateqq_combined}
\end{figure}
For easier comparison of RMSEs between IS, TIS and PSIS,
Figure~\ref{fig:exptest_relrmse}  shows the relative RMSEs.

Figure~\ref{fig:exptest_convrateqq_combined} shows how the relative convergence
rate depends on $h$-specific $\hat{k}_h$ estimates and matches the theoretical results in Appendix \ref{sec:truncated_mean}. For
PSIS we observe convergence rates as discussed in Section \ref{sec:PSIS_RMSE} and Appendix \ref{sec:truncated_mean}. The shown $\hat{k}_h$ are estimated
using the mean from 1000 simulations, with $S=10^5$ for each simulation.
Empirical convergence rates are estimated by a linear fit to the RMSE results illustrated
in Figures~\ref{fig:exptest_error0}--\ref{fig:exptest_error2}. Previously in the literature, the focus has been mostly
on the binary decision between $k<0.5$ and $k \geq 0.5$. As discussed in Section \ref{sec:rate} and Appendix \ref{sec:rate_appendix}, with finite sample size the PSIS convergence
rate starts to differ from CLT convergence rate approximately when $k>0.5-0.5/\log_{10}(S)$. In this and the other experiments in the paper, the convergence
rate at $k=0.5$ matches well the theoretically predicted $1-1/\log(S)$, for example $0.89$ when $S=2000$. Correspondingly at $k=0.7$, the convergence rate is close to $\alpha\approx 0.6$ which matches the theoretical $2-2k$. In this
experiment we added results for $\lambda \in (1.9, 2.1)$ which have $k$ just
below and above $0.5$, illustrating that there is no sharp transition. A dashed
line has been drawn from $k=0.5, \alpha=1$ to $k=0, \alpha=0$ to illustrate the asymptotic behavior when $S\rightarrow\infty$.
Based on analytic consideration, we assume that $h$ specific $\hat{k}_h$ (as described in Section \ref{s:pareto}) for $\theta r(\theta)$ and $\theta^2 r(\theta)$ (first and second moment) are not as accurate
as $\hat{k}$ estimates for $r(\theta)$  (zeroth moment). This is likely
explanation why the corresponding empirical convergence rate curves go slightly
beyond the theoretical line.
PSISa has better relative convergence rate due to additional variance reduction by using $M=S$, but the difference is small and using $M=S$ is not usually applicable as the whole distribution is not Pareto distribution.

\subsection{Univariate Normal and Student's $t$}
\label{sec:norm_and_t_test}

The previous example with exponential target and proposal is especially suited
for PSIS, as the whole importance ratio distribution is well fitted with the
generalized Pareto distribution. In this section, we consider various univariate
target and proposal distribution combinations to show the behavior in the case
of different target-proposal tail combinations. In the next section we examine
the corresponding multivariate cases.

We do not use the simulated example by \citet{Ionides:2008} having the target
$p(\theta)=\mbox{normal}(\theta\,|\,0,1)$ and the proposal
$g(\theta)=\mbox{normal}(\theta\,|\,0,\sigma)$, as both distributions have the same mean and
thus when estimating the first moment the lowest RMSE would be obtained by using
identity weights and $\sigma\rightarrow 0$.

To test the performance for the zeroth, first and second moments, we choose the proposal
distributions to have different mean and scale than the target distribution. The
tested pairs are
\begin{enumerate}
\item $p(\theta)=\mbox{normal}(0,1)$, $g(\theta)=\mbox{normal}(\mu, 0.8)$:
 This is a special case with the matching tail shape of the target and proposal
 leading to a case which is favorable for PSIS.
\item $p(\theta)=t_{20}(0,1)$, $g(\theta)=\mbox{normal}(\mu, 0.9)$:
 This resembles an applied case of using a Gaussian posterior approximation when
 the target has a thicker tail.
\item $p(\theta)=t_{20}(0,1)$, $g(\theta)=t_{21}(\mu, 0.8)$:
 This resembles an applied case of leave-one-out importance sampling where the
 proposal has a slightly thinner tail.
\item $p(\theta)=t_{7}(0,1)$, $g(\theta)=t_{8}(\mu, 0.8)$: This resembles an
 applied case of leave-one-out importance sampling where the proposal has a slightly thinner tail, but both having thicker tails than in case 3.
\end{enumerate}
Here we have left out easy univariate examples where the proposal would have a thicker tail
than the target. To vary how well the proposal matches the target, the mean of
the proposal $\mu$ is varied.

\begin{figure}[tp]
  \begin{center}
    \includegraphics[width=\textwidth]{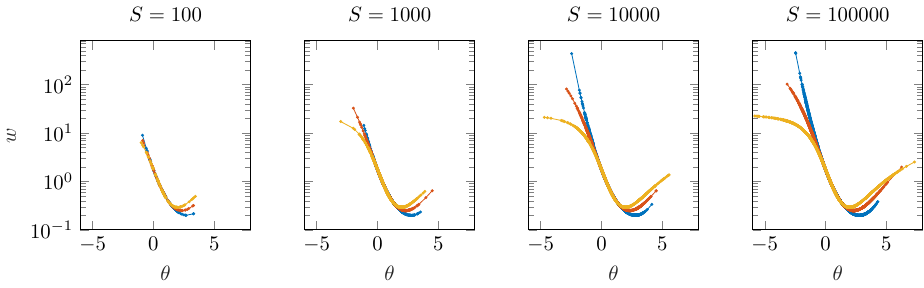}
  \end{center}
  \vspace{-.2in}
  \caption{Plain IS weights plotted by $\theta$ for different
  univariate target-proposal pairs:  {\color{matblue}$p(\theta)=\mbox{normal}(0,1)$, $g(\theta)=\mbox{normal}(1.5, 0.8)$ (blue)},
  {\color{matred}$p(\theta)=t_{20}(0,1)$, $g(\theta)=t_{21}(1.5, 0.8)$ (red)},
  {\color{matyellow}$p(\theta)=t_{7}(0,1)$, $g(\theta)=t_{8}(1.5, 0.8)$ (yellow)}, and different
  sample sizes $S$.}
  \label{fig:xoff_xvsweight}
\end{figure}

\begin{figure}[tp]
  \begin{center}
    \includegraphics[width=\textwidth]{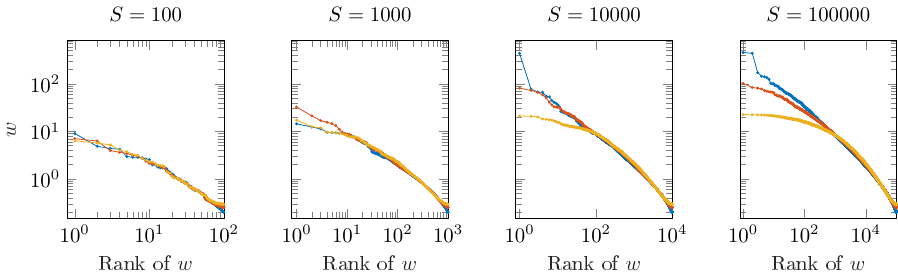}
  \end{center}
  \vspace{-.2in}
  \caption{Sorted plain IS weights plotted by rank for different
  univariate target-proposal pairs:  {\color{matblue}$p(\theta)=\mbox{normal}(0,1)$, $g(\theta)=\mbox{normal}(1.5, 0.8)$ (blue)},
  {\color{matred}$p(\theta)=t_{20}(0,1)$, $g(\theta)=t_{21}(1.5, 0.8)$ (red)},
  {\color{matyellow}$p(\theta)=t_{7}(0,1)$, $g(\theta)=t_{8}(1.5, 0.8)$ (yellow)}, and different
  sample sizes $S$.
  }
  \label{fig:xoff_weightsb}
\end{figure}

Figure~\ref{fig:xoff_xvsweight} shows plain importance sampling weights for different $\theta$,
different target-proposal pairs (1, 3 and 4 in the above list with $\mu=1.5$),
and different sample sizes $S$. With increasing sample size $S$ we get more
draws from the tails and the differences between the weight functions become
more apparent. To better illustrate how the tail shape of the empirical weight
distributions change when the sample size $S$ increases,
Figure~\ref{fig:xoff_weightsb} shows the same weights sorted and plotted by
rank.
With $S=100$ the weight distributions look similar and the corresponding
$\hat{k}$'s are $0.66$, $0.66$, and $0.64$. In the last case the weight distribution is bounded, but $S=100$ is not sufficient to observe the bound with non-negligible probability. As $S$ increases the distributions
of the weights eventually look different and the corresponding $\hat{k}$'s are
$0.66$, $0.47$, and $-0.28$. In the bounded ratio distribution case, we are observing the bound. This shows that using only a small portion of the
raw weights in the tail allows the $\hat{k}$ diagnostic to adapt to the
empirically observed tail shape, but also that any finite sample result can be different from a result with much bigger sample size. Figure~\ref{fig:xoff_weightsb} also illustrates
the motivation to truncate the weights at the maximum raw weight value
$\max(r_s)$. This will allow the use of larger $M$ for the Pareto fit to reduce
the variance, while being able to adapt when the magnitude of the weights is
saturating with increasing $S$.

\begin{figure}[tp]
  \begin{center}
    \includegraphics[width=\textwidth]{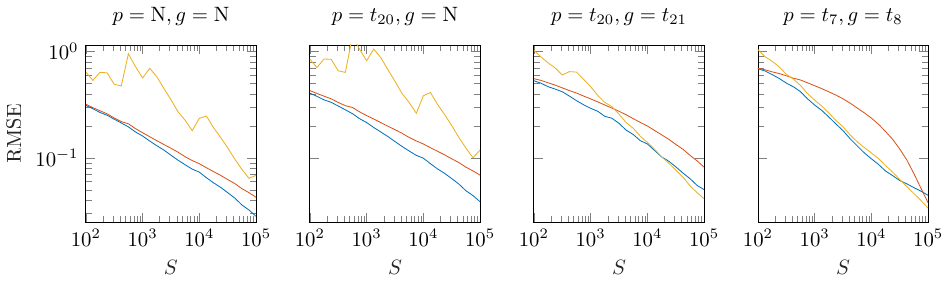}
  \end{center}
  \vspace{-.2in}
  \caption{{\color{matyellow}IS is yellow}, {\color{matred}TIS is red}, and {\color{matblue}PSIS is blue}.
    The mean RMSE from 1000 simulations for the different univariate target-proposal pairs
with $\mu$ from left to right being $1.5$, $2$, $2.25$, and $3$. }
  \label{fig:xoff_rmsevsproposal}
\end{figure}

Based on the above $\hat{k}$ values we may assume that PSIS is beneficial for
small $S$, and for cases 3 and 4 when $S$ grows eventually plain IS will also
work well. Figure~\ref{fig:xoff_rmsevsproposal} shows the mean RMSE from 1000
simulations for the four different target-proposal pairs as listed above with
hand-picked $\mu$ values to illustrate the typical behavior of IS, TIS and PSIS
when $\hat{k} \in(0.5,0.7)$ (for very low $\hat{k}$ the differences are negligible
and for very high values all methods fail). PSIS is able to adapt well in all
cases and has the smallest RMSE in almost all cases. When the proposal
distribution has a thin tail (e.g., Gaussian in cases 1 and 2), IS has high
variance and the variance remains high with increasing $S$.
If the tail of the proposal is thick (e.g., Student's $t$) and asymptotically the
weight distribution has a short tail (small $k$), PSIS performs better than IS
for small $S$. Eventually, the RMSE of IS can get as small as the RMSE of PSIS.
Most of the time, TIS has a larger RMSE than PSIS. For thick tailed proposals, TIS
is not able to adapt well to the changing shape of the empirical weight
distribution.

\begin{figure}[tp]
  \begin{center}
    \includegraphics[width=\textwidth]{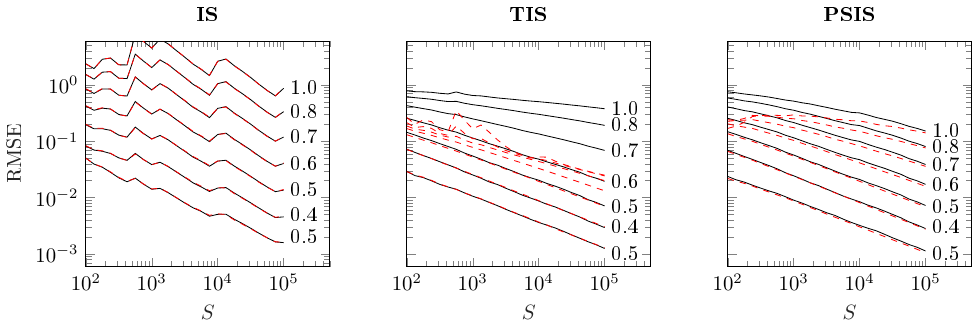}
  \end{center}
  \vspace{-.2in}
  \caption{RMSE (black lines) and {\color{red}the mean of the MCSE estimates (red
  lines)} for the zeroth moment estimate. The target distribution is univariate $t_{20}(0,1)$
  and the proposal distribution is univariate $\mbox{normal}(\mu,0.9)$, with $\mu \in (0, 0.5, 1.0,
  1.5, 2.0, 2.5, 3.0)$. For each graph, the lines are ordered with low values of
  theta at bottom and high values at top, with high $\theta$ values leading to
  high RMSE and high $\hat{k}$. The numbers at the end of black lines are
  average $\hat{k}$ values estimated when $S=10^5$.}
  \label{fig:xoffnt20s_error0}
\end{figure}

Figure~\ref{fig:xoffnt20s_error0} shows as a representative example of the RMSE
and MCSE estimates for zeroth moment estimated with IS, TIS and PSIS
in case of $p(\theta)=t_{20}(0,1)$, $g(\theta)=N(\mu, 0.9)$. PSIS is more
stable, has smaller RMSE than IS, and has more accurate MCSE
estimates than TIS.

\begin{figure}[tp]
  \begin{center}
    \includegraphics[width=\textwidth]{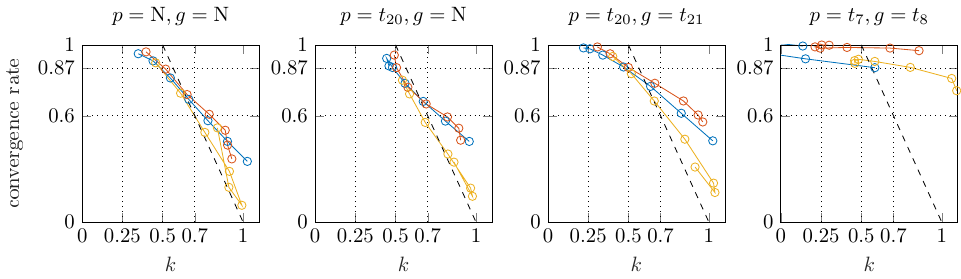}
  \end{center}
  \vspace{-.2in}
  \caption{Practical convergence rates for PSIS  estimating {\color{matblue}zeroth
  (blue \textbf{--})}, {\color{matred}first (red \textbf{--})} and {\color{matyellow}second (yellow \textbf{--})} moments. Different univariate target-proposal distribution
  pairs are used in different subplots. $\hat{k}$ estimates in case of first and
  second moments are $h$ specific. Dotted horizontal lines at $0.87$ and $0.6$ show
  the typical convergence rates at $\hat{k}=0.5$ and $\hat{k}=0.7$ for several
  different experiments in this paper.}
  \label{fig:xoff_convrateqq_combined}
\end{figure}

Figure~\ref{fig:xoff_convrateqq_combined} shows the relative convergence rate
with respect to $h$-specific $\hat{k}_h$ estimates. We observe similar behavior as
in the exponential distribution example, even now the ratio distributions are not exactly Pareto shaped. For thick tail proposal
distributions we tend to overestimate $\hat{k}_h$ values more than for the thin
tailed proposal distributions, but the convergence rates stay good. Given
estimated $\hat{k}_h$ values, we can use the theoretical result to provide a conservative convergence rate
estimate (it is better that the diagnostic is conservative than too optimistic).

\subsection{Multivariate Normal and Student's $t$}

In this section, we consider the isotropic multivariate versions of the four
target-proposal pairs as the number of dimensions increases. In addition, we
also examine a case where the proposal has thicker tails than the target, and
show that with an increasing number of dimensions this will also lead to
increasing variability of the weights. The compared target-proposal pairs are
\begin{enumerate}
\item $p(\theta)=\mbox{MVN}(\mathbf{0},I)$, $g(\theta)=\mbox{MVN}(0.4 \cdot \mathbf{1}, 0.8 I)$
\item $p(\theta)=t_{20}(\mathbf{0},I)$, $g(\theta)=\mbox{MVN}(0.4 \cdot \mathbf{1}, 0.9 I)$
\item $p(\theta)=t_{20}(\mathbf{0},I)$, $g(\theta)=t_{21}(0.4 \cdot \mathbf{1}, 0.8 I)$
\item $p(\theta)=t_{7}(\mathbf{0},I)$, $g(\theta)=t_{8}(0.4 \cdot \mathbf{1}, 0.8 I)$
\item $p(\theta)=\mbox{MVN}(\mathbf{0},I)$, $g(\theta)=t_{7}(0.4 \cdot \mathbf{1}, 0.8 I)$
\end{enumerate}
In all these cases, the proposal distribution is just slightly displaced and
with slightly narrower scale. The fifth proposal distribution has thicker tails
than the target distribution, so that the importance ratios are bounded, although that bound can be too far to be observed with a finite sample size. The number of dimensions is varied as
$D \in (1, 2, 4, 8, 16, 32, 64)$.

\begin{figure}[tp]
  \begin{center}
    \includegraphics[width=\textwidth]{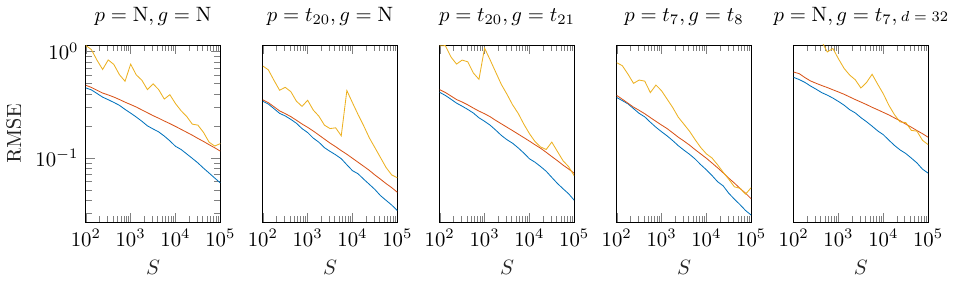}
  \end{center}
  \vspace{-.2in}
  \caption{{\color{matyellow}IS
  is yellow \textbf{---}}, {\color{matred}TIS is red \textbf{---}}, and {\color{matblue}PSIS is blue \textbf{---}}. The mean RMSE from 1000 simulations for the different multivariate target-proposal pairs
with $D=16$ for all except the last one with $D=32$ (the number of dimensions for each plot were chosen to show interesting behavior between too easy and too difficult).}
  \label{fig:dimxoff_rmsevsproposal}
\end{figure}

Figure~\ref{fig:dimxoff_rmsevsproposal} shows the mean RMSE from 1000
simulations for the five different target-proposal pairs as listed above with
$D=16$ for all except the last case with $D=32$. The plots illustrate the
typical behavior of IS, TIS and PSIS when $0.5 < \hat{k} < 0.7$. PSIS has the smallest RMSE in all cases. TIS has a slower convergence
rate than PSIS. IS is overall more unstable and has higher RMSE. Comparing the
case with a thick tailed proposal distribution to the corresponding univariate
case, we see that IS requires larger $S$ before beginning to stabilize. The
rightmost subplot illustrates that even if the proposal distribution has thicker
tails than the target distribution and the importance ratios are bounded, the
variance of the importance ratio distribution can be high when the number of
dimensions $D$, and IS is not a safe choice in finite sample case.

\begin{figure}[tp]
  \begin{center}
    \includegraphics[width=\textwidth]{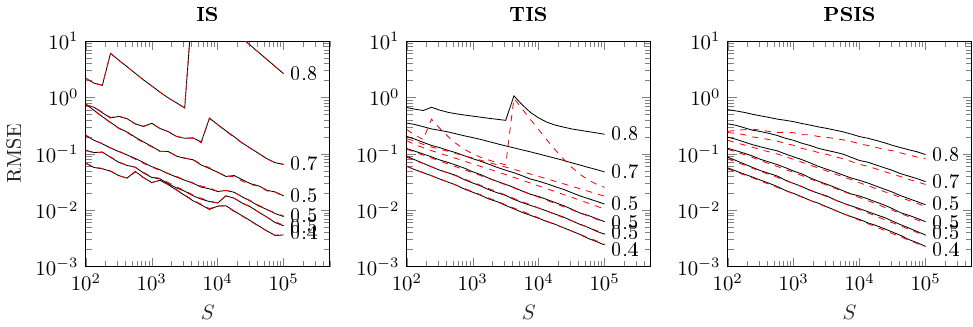}
  \end{center}
  \vspace{-.2in}
  \caption{RMSE (black line) and {\color{red}the mean of the MCSE estimates (red
  dashed line)} for the zeroth moment estimate. The target distribution is
  multivariate $t_{20}(\mathbf{0},I)$ and the proposal distribution is multivariate $\mbox{MVN}(0.4 \mathbf{1},
  0.9 I)$, with the number of dimensions $D \in (1, 2, 4, 8, 16, 32, 64)$. For
  each graph, the lines are ordered with low values of $D$ at bottom and high
  values at top, with high $D$ values leading to high RMSE and high $\hat{k}$.
  The numbers at the end of black lines are average $\hat{k}$ values
  estimated when $S=10^5$.}
  \label{fig:dimxoffnt20s_error0}
\end{figure}

Figure~\ref{fig:dimxoffnt20s_error0} shows a representative example of the RMSE
and Monte Carlo estimates for zeroth moment estimated with IS, TIS and PSIS with
$p(\theta)=t_{20}(\mathbf{0},I)$, $g(\theta)=N(0.4 \cdot \mathbf{1}, 0.9 I)$.
PSIS is more stable, has smaller RMSE than IS and TIS, and has more accurate
MCSE estimates than TIS. All methods eventually fail as the number of
dimensions increases and even a small difference in the distributions is
amplified. Having a bounded importance ratios and finite variance, does not
prevent the behavior to be indistinguishable from the infinite variance case in finite sample case.
In these multivariate examples, we observe sudden large jumps also for TIS.
Truncation in TIS  fails when there is one extremely large weight that causes
the truncation level to rise so high that other large weights are not truncated.
PSIS performs better in the same situation, as one extreme large weight doesn't
affect the generalized Pareto fit as much.

\begin{figure}[tp]
  \begin{center}
    \includegraphics[width=\textwidth]{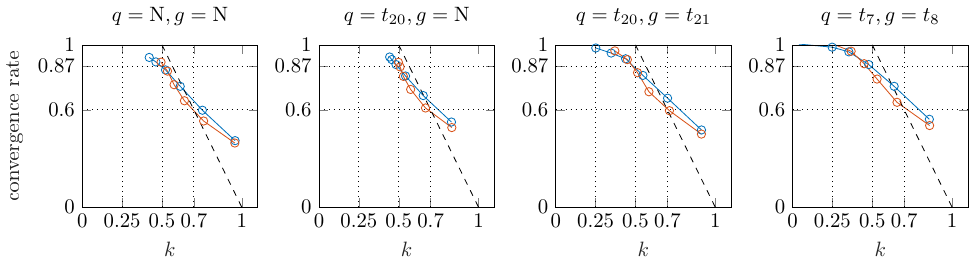}
  \end{center}
  \vspace{-.2in}
  \caption{Practical convergence rates for PSIS  estimating {\color{matblue}zeroth
  (blue}, and {\color{matred}first (red)} moments. Different multivariate target-proposal distribution
  pairs are used in different subplots. $\hat{k}$ estimates in case of first and
  second moments are $h$ specific. Dotted horizontal lines at $0.87$ and $0.6$ show
  the typical convergence rates at $\hat{k}=0.5$ and $\hat{k}=0.7$ for several
  different experiments in this paper.} \label{fig:dimxoff_convrateqq_combined}
\end{figure}

Figure~\ref{fig:dimxoff_convrateqq_combined} shows the practical convergence
rate with respect to the $h$-specific $\hat{k}$ estimates. We observe similar
behavior as in the univariate example. As we saw before, the observed
convergence rate with respect to $\hat{k}$ is close to the theoretical result.

\section{$h$-specific $\hat{k}_h$}

In Section \ref{sec:k_hat_big} we defined $h$-specific $\hat{k}_h$ as
maximum of $\hat{k}$ estimated with $M$ smallest and $M$ largest
values of $h(\theta)r(\theta)$ (left and right tail). As we choose the
larger value from two estimated values with uncertainty, $\hat{k}_h$
has positive bias when those values are close to each other. We think
this bias is acceptable as a conservative choice. The benefits of this
approach are that it is simple and it is also natural when assessing
tail shape of $h(\theta)$ given Monte Carlo draws (i.e.,
$r(\theta)=1$).

In an earlier version of this paper we proposed to estimate
$h$-specific $\hat{k}_h$ for $\sqrt{1+h(\theta)^2}\,r(\theta)$, a
form used by \citet{Epifani+MacEachern+Peruggia:2008} to avoid
zero values when proving conditions for CLT for different IS-LOO
influence measures. This form is also close to 
$|h(\theta)|r(\theta)$, which has been commonly used to evaluate
$h$-specific variance of IS estimators \citep[e.g.,][Ch. 9]{Owen:2013}. The benefit is that we could estimate just
one $\hat{k}_h$, but the downside is that the tail can then be a
mixture of two tails, and in a finite sample case this can lead to
underestimating $k_h$.  In our experiments the two tail diagnostic of
$h(\theta)r(\theta)$ worked slightly better over different scenarios.

Although we recommend use of $h$-specific $\hat{k}_h$, we do not
recommend smoothing $h(\theta)r(\theta)$. Smoothing $r(\theta)$ and
$h(\theta)r(\theta)$ separately introduces an additional bias in the
normalization, which based on our experiments leads in some cases to
noticeable increase in the estimation error.

\section{Marginal Distribution of $k$}
\label{sec:marginal_k}

The method by \citet{Zhang+Stephens:2009} uses profile likelihood,
weakly informative empirical data based prior, and quadrature
integration. \citet{Zhang+Stephens:2009} analyse the derived
posterior mean estimate $\hat{k}$, for which they report only a small
bias. In this paper we focused on using $\hat{k}$ as the diagnostic,
as we didn't find additional benefit from estimating, for example, the
probability that $k<0.5$ or that $k<0.7$. However, knowing the
associated uncertainty can be sometimes useful to assess what could be
learned by increasing the sample size in importance sampling.

\begin{figure}[t]
  \begin{subfigure}{0.31\textwidth}
    \vspace{16.5mm}
    \includegraphics[width=\textwidth]{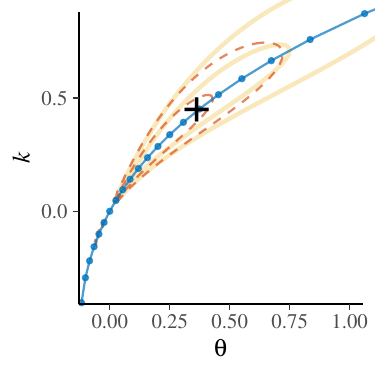}
    \vspace{-8mm}
    \caption{}
  \end{subfigure}
  ~
  \begin{subfigure}{0.67\textwidth}
    \includegraphics[width=\textwidth]{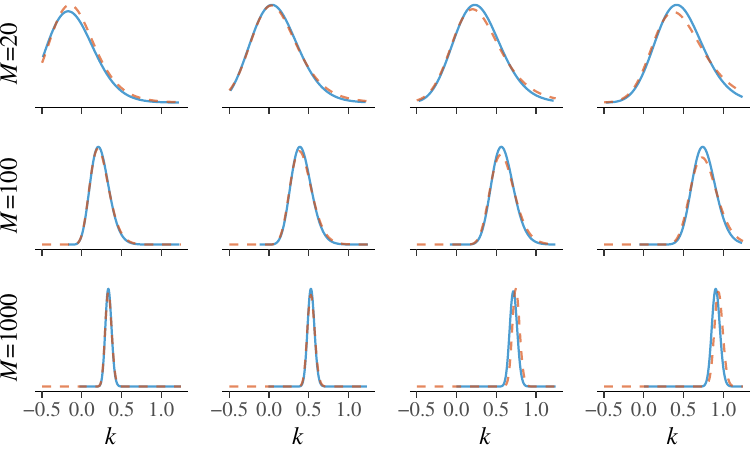}
    \vspace{-7mm}
    \caption{}
  \end{subfigure}
  \vspace{-0.5\baselineskip}
  \caption{(a) {\color{matyellow} The joint likelihood}, {\color{matred} the joint posterior}, {\color{matblue} the profile curve and the quadrature points}, and the profile posterior mean $(\hat{\theta},\hat{k})$ marked with $\boldsymbol{+}$, based on 20 observations simulated from the generalized Pareto distribution (GPD) with $k=0.7$ and $\sigma=1$. (b) Posterior marginal of $k$ estimated from {\color{matblue} the profile posterior}, and {\color{matred} exact posterior marginal (dashed line)}. Rows show example marginals for simulated data with sizes of 20, 100, and 1000. Columns show example marginals for simulated data generated from GPD with $k \in (0.3,0.5,0.7,0.9)$. The third subplot on the top row is for the same data as in the plot (a).}
\label{fig:gpd_marginal}
\end{figure}

\citet{Zhang+Stephens:2009} use parameterization $(\theta,k)$, where
$\theta=\sigma/k$. Figure~\ref{fig:gpd_marginal}a shows the joint
likelihood, joint posterior, and the profile curve and the quadrature
points. The profile likelihood for $\theta$ is the likelihood along a
curve with $k$ chosen to maximize the likelihood. When the prior is set
only on $\theta$, the profile posterior is determined along the same
curve. \citet{Zhang+Stephens:2009} select the location of the
quadrature points along the profile curve and spacing is determined by
the prior. Then the quadrature weights to determine the profile
posterior mean are simply the normalized likelihood values at the
quadrature points. After finding the posterior mean $\hat{\theta}$, the
estimate $\hat{k}$ comes from the profile curve, and
$\hat{\sigma}=\hat{\theta}/\hat{k}$.

\citet{Zhang+Stephens:2009} discuss only point estimates $\hat{k}$ and
$\hat{\sigma}$, but we demonstrate that the profile posterior can be
used as an approximation for the marginal posterior of $k$. When
computing the posterior mean with quadrature, the quadrature weights
are proportional to the likelihood as the quadrature points are spaced
according to the prior density. To plot the profile posterior, we
compute the normalized posterior densities at the quadrature points.
Figure~\ref{fig:gpd_marginal}b compares the profile posterior and true
marginal posterior (computed from the exact two dimensional posterior
using adaptive quadrature integration).  Figure
\ref{fig:gpd_marginal}b also illustrates the
uncertainty with different values of $M = (20,100,1000)$, which
correspond to tail sample sizes using the proposed rule for total
sample sizes $S = (100,1111,111\,111)$.

\citet{Zhang:2010:improving} presented an alternative empirical prior
which in case of small sample sizes produces estimates with smaller
bias for estimating $\sigma$ and $k$ when $k>0.5$, but has slightly
lower efficiency for estimating $k$ when $0<k<4$. Compared to the
uncertainty in $k$ (see Figure \ref{fig:gpd_marginal}b), the small
differences in bias and efficiency \citep[as reported by][]{Zhang:2010:improving} when
$k\in(0,1)$ do not make a practical difference in PSIS or diagnostics.

\section{Regularization of Pareto Fit for Small $S$}
\label{sec:regularization_of_k}

To reduce the variance of the $\hat{k}$ estimate for small $S$ ($S$
less than about 1000), we use an additional regularization,
\begin{align*}
\hat{k}=(M\hat{k}+10\cdot 0.5)/(M+10),
\end{align*}
which corresponds approximately to a weakly informative Gaussian prior
that has a weight of 10 observations from the tail shrinking toward
$0.5$.
If $S=100$ and using the proposed rule to choose $M=20$, we see that
$\hat{k}$ is strongly regularized. If $S=1000$ and using the proposed
rule to choose $M=95$, we see that the regularization has only weight
of one tenth of the data.
This regularization adds bias toward $0.5$, but
based on our simulation results, this reduced the variance and RMSE of
the PSIS estimates with small $S$, without introducing significant
bias or affecting the RMSE for larger $S$. In many applications we
would assume $S>1000$, but in sequential and iterative methods such as
particle filters and black box variational inference we may want to
use small $S$ for computational speed issues.

\section{Adjusting $M$ in Case of Dependent MCMC Draws}
\label{sec:adjusting_M_MCMC}

We adjust the algorithm in the following way to take into account the
usually smaller effective sample size of dependent MCMC draws.
Due to correlated MCMC draws, we use more ratios in the tail to keep
the variance of $\hat{k}$ in control.  We adjust the number of tail
ratios used as $M = \lfloor \min(0.2 S, 3\sqrt{S/R_{\effmcmc}}) \rfloor$,
where $R_{\effmcmc}=S_{\effmcmc}/S$ is the estimated relative efficiency of the
MCMC draws, and the effective sample size $S_{\effmcmc}$ for $r(\theta)$ is
computed using the split-chain effective sample size estimate method
\citep{Vehtari+etal:2021:Rhat}. This effective sample size is
not directly the effective sample size for $\hat{k}$, but it is
related and much easier to compute.

\section{Stan Program for Linear Regression on the Stack Loss Data}
\label{sec:stan_stack_code}

\begin{small}\begin{quotation}\noindent\vspace{-1.5\baselineskip}
\begin{Verbatim}
data {
  int<lower=0> N;
  int<lower=0> p;
  vector[N] y;
  matrix[N,p] x;
}
transformed data { // to standardize the x's
  matrix[N,p] z;
  vector[p] mean_x;
  vector[p] sd_x;
  real sd_y = sd(y);
  for (j in 1:p) {
    mean_x[j] = mean(col(x,j));
    sd_x[j] = sd(col(x,j));
    for (i in 1:N) z[,j] = (x[i,j] - mean_x[j]) / sd_x[j];
  }
}
parameters {
  real beta0;
  vector[p] beta;
  real<lower=0> sigmasq;
  real<lower=0> phi;
}
transformed parameters {
  real<lower=0> sigma;
  vector[N] mu;
  sigma = sqrt(sigmasq);
  mu = beta0 + z * beta;
}
model {
  beta0 ~ normal(0, 100);
  phi ~ cauchy(0, sd_y);
  beta ~ normal(0, phi);
  sigmasq ~ inv_gamma(0.1, 0.1);
  y ~ normal(mu, sigma);
}
generated quantities {
  vector[N] log_lik;
  for (i in 1:N) log_lik[i] = normal_lpdf(y[i] | mu[i], sigma);
}
\end{Verbatim}
\end{quotation}
\end{small}

\bibliography{psis}

\end{document}